\documentclass[article]{elsarticle}
\usepackage{todonotes}
\usepackage{lineno,hyperref,microtype,amssymb,amsmath}
\usepackage{color}
\modulolinenumbers[1]

\makeatletter
\def\ps@pprintTitle{%
	\let\@oddhead\@empty
	\let\@evenhead\@empty
	\def\@oddfoot{\centerline{\thepage}}%
	\let\@evenfoot\@oddfoot}

\usepackage{color} 

\begin{document}

\begin{frontmatter}

\title{A generalised framework for detailed classification of swimming paths inside the Morris Water Maze}

\author[mymainaddress,email1]{Avgoustinos Vouros}
\author[mymainaddress]{Tiago V. Gehring}
\author[mymainaddress2]{Kinga Szydlowska}
\author[mymainaddress2]{Artur Janusz}
\author[mymainaddress]{Mike Croucher}
\author[mymainaddress2]{Katarzyna Lukasiuk}
\author[mymainaddress2]{Witold Konopka}
\author[email3]{Carmen Sandi}
\author[mymainaddress]{Zehai Tu}
\author[mymainaddress,email2]{Eleni Vasilaki}



\address[mymainaddress]{The University of Sheffield, Faculty of Engineering, Department of Computer Science, Sheffield, UK}
\address[mymainaddress2]{Nencki Institute of Experimental Biology, Department of Molecular and Cellular Neurobiology, Warsaw, Poland}
\address[email3]{Laboratory of Behavioral Genetics, Brain Mind Institute, EPFL, Lausanne, Switzerland}
\address[email1]{avouros1@sheffield.ac.uk}
\address[email2]{e.vasilaki@sheffield.ac.uk}

\begin{abstract}
The Morris Water Maze is commonly used in behavioural neuroscience for the study of spatial learning with rodents. Over the years, various methods of analysing rodent data collected during this task have been proposed. These methods span from classical performance measurements (e.g. escape latency, rodent speed, quadrant preference) to more sophisticated methods of categorisation which classify the animal swimming path into behavioural classes known as exploration strategies. Classification techniques provide additional insight about the different types of animal behaviours but still only a limited amount of studies utilise them mainly because they highly depend on machine learning knowledge. We have previously demonstrated that the animals implement various strategies and by classifying whole trajectories can lead to the loss of important information. In this work, we developed a generalised and robust classification methodology which implements majority voting to boost the classification performance and successfully nullify the need of manual tuning. Based on this framework, we built a complete software, capable of performing the full analysis described in this paper. The software provides an easy to use graphical user interface (GUI) through which users can enter their trajectory data, segment and label them and finally generate reports and figures of the results.
\end{abstract}

\begin{keyword}
Morris Water Maze \sep navigation \sep behavioural category \sep semi-supervised clustering \sep learning memory \sep classification boosting
\end{keyword}

\end{frontmatter}


\section{Introduction}



The Morris Water Maze (MWM), designed by Richard Morris, was first described back in 1981 in a study regarding the spatial localisation of rats \cite{morris1981spatial}. The MWM quickly became very popular and by the end of the eighties a large number of published work using the MWM had been reported \cite{brandeis1989use}. For instance, the review work of D'Hooge and Deyn mentions more than 2000 publications related to the MWM task within the decade 1990-2001 \cite{d2001applications}. More recently, virtual forms of the MWM have been used directly on human subjects and this generalisation made it possible to comparatively assess human and rodent place navigation \cite{schoenfeld2017variations}, compare spatial learning between sexes \cite{astur1998characterization} and directly study how certain factors (e.g. stimuli, age, etc.) affect the spatial navigation and how certain areas of the brain perform under the effects of them (\cite{cornwell2008human},\cite{daugherty2015changes},\cite{piber2016mineralocorticoid},\cite{korthauer2017cognitive}).

In a typical MWM experiment the rodent is placed inside a circular pool filled with water and is tasked to find a hidden platform, which is placed in one of the four quadrants of the pool. Since the animal is unable to see the platform, it has to rely on external visual cues in order to navigate inside the pool and find the platform. After a number of trials, it is expected that the animal will have learned the location of the platform and will therefore be able to find it in less time than in the beginning of the trials \cite{morris1984developments}.

Most of the studies using the MWM experiment utilise several measurements of performance in order to assess learning and memory. Many of these measurements have also been used to ensure that the animal groups have equal skills and abilities (e.g. swimming ability, speed, `understanding' of the escape mechanism) \cite{vorhees2014assessing,maei2009most}. Notable measurements include the time that the animal spends inside each quadrant of the pool, the latency of finding the platform in each trial, the directionality and the total swimming distance on each trial \cite{morris1984developments,brandeis1989use,lindner1997reliability}. There are also a number of more sophisticated measurements such the body temperature of the animals throughout the experiment \cite{lindner1991relationship} or the cumulative distance to platform, which is the distance between the animal location and the platform location calculated a number of times with a specific sampling rate \cite{gallagher1993severity,dalm2000quantification}.

These simplistic measurements and statistics have been criticised as insufficient to capture all the different animal behaviours that are present during the MWM experiments \cite{dalm2000quantification,gehring2015detailed}. For this reason researchers started to study the various behaviours that the animals were expressing inside the pool, which are known as exploration strategies. Notable are the studies of Wolfer et al., who computed a large amount of measures for each animal swimming path inside the maze in order to categorise the various animal strategies \cite{wolfer2000dissecting,wolfer2001extended,wolfer1998spatial}. Other studies include the automatic classification procedures of Graziano et al. \cite{graziano2003automatic} and Garthe et al. \cite{garthe2009adult}. Both of them specified regions of interest inside the arena but the categorisation method of Graziano et al. was based on a number of path measures while in work of Garthe et al. a hierarchical classification algorithm was used and the categorisation of each swimming path was mainly based on the amount of time that the animal spent in each region. The latter method was also used in more recent studies (\cite{rogers2017search},\cite{yeshurun2017elevated}).

A point of criticism of the aforementioned studies is that the classification procedure requires prior knowledge of animal behaviour strategies, while manual classification of each trajectory is subject to bias \cite{illouz2016unraveling}. Thus the study of Illouz et al. attempted to minimise the human error and to create an unbiased analysis by an automatic classification procedure based on support vector machines \cite{cortes1995support}. In their work, they extracted a set of 11 features from the X and Y coordinates of the animal trajectories and using a database of labelled data (swimming paths with pre-defined classes) they were able to automatically classify a large amount of full animal trajectories by performing a series of hierarchical decisions.

In our previous work \cite{gehring2015detailed} we argued that animals employ several strategies during each trial in order to find the platform and by assigning whole animal trajectories to single behavioural classes results in the loss of important information. For this reason, we proposed a more sophisticated automatic quantification methodology capable of classifying and presenting the various animal behaviours in much more detail during each trial. According to this approach, the animal swimming path is first split into segments and then the segments are classified into behavioural strategies. In this way, changes in the animal behaviour within each trial can be detected and the animal swimming path, as a whole, falls under more than one strategies, revealing how the animal behaviour involved within the trial.  

For the classification of the segments we used a semi-supervised classification procedure which requires manual classification (labelling) of a small amount of data. An advantage of this procedure was that our classification is based on a clustering algorithm which is able to detect patterns in the data. Therefore, the behavioural classes didn't necessarily have to be defined a priori. On the other hand, the method developed in our previous study required a certain degree of knowledge about machine learning methods, which prevented the direct application of the methodology to other datasets.

In this work, we present an automatic boosted classification procedure based on majority voting, which improves on the classification error, and a validation framework which leads to conclusions with a high degree of confidence. Finally, the software tools implemented for our previous work \cite{gehring2015detailed} have been re-engineered in order to produce a fully working software capable of performing all of our analyses, without requiring machine learning knowledge from the user. This software is called RODA (ROdent Data Analytics) \cite{avgoustinos_vouros_2017_1117837} and is focused on the MWM experiment. It provides an easy to use graphical user interface (GUI) for loading the data and defining the experimental specifications. It also supports automatic segmentation and semi-automatic classification, and produces quality figures which can be exported into various image formats. The software is available on the  github repository \href{https://github.com/RodentDataAnalytics/mwm-ml-gen}{https://github.com/RodentDataAnalytics/mwm-ml-gen} under the GNU General Public License version 3 (GPL-3.0).

\section{Materials and Methods}

\subsection{Analysis Overview}
In our proposed analysis method, the swimming paths of the animals inside the Morris Water maze are divided into segments of approximately equal length and a certain overlap percentage. For each segment a set of eight features is computed. The features are then used in the classification procedure. Finally, a small portion of the segments needs also to be assigned manually to a specific strategy (labelling); this information is used as prior knowledge to guide the classification procedure.  

Our classification procedure, which assigns segments to classes of behaviour, is based on a semi-supervised clustering algorithm called Metric Pairwise Constrained K-Means (MPCK-Means) \cite{bilenko2004integrating}. This algorithm incorporates the two main approaches of semi-supervised clustering: metric learning (the measuring of similarity, `distance', between data) and constrained-based learning (the use of labels or constrains that produce a better grouping of the data) \cite{bilenko2004integrating}. To turn the algorithm into a classifier, the labelled data were used not only to guide the clustering process but also to assign clusters to classes (see supplementary material for more details). 

However, a common issue with many clustering algorithms, including MPCK-Means, is that a predefined number of target clusters needs to be provided; this number indicates the amount of clusters in which our data will be partitioned. Determining the optimal number of target clusters is challenging and, although many different quality measures were proposed over time \cite{kovacs2005cluster}, this value will depend on the specific clustering method and data at hand. 

\begin{figure}[h!]
	\begin{center}
		\includegraphics[width=\textwidth]{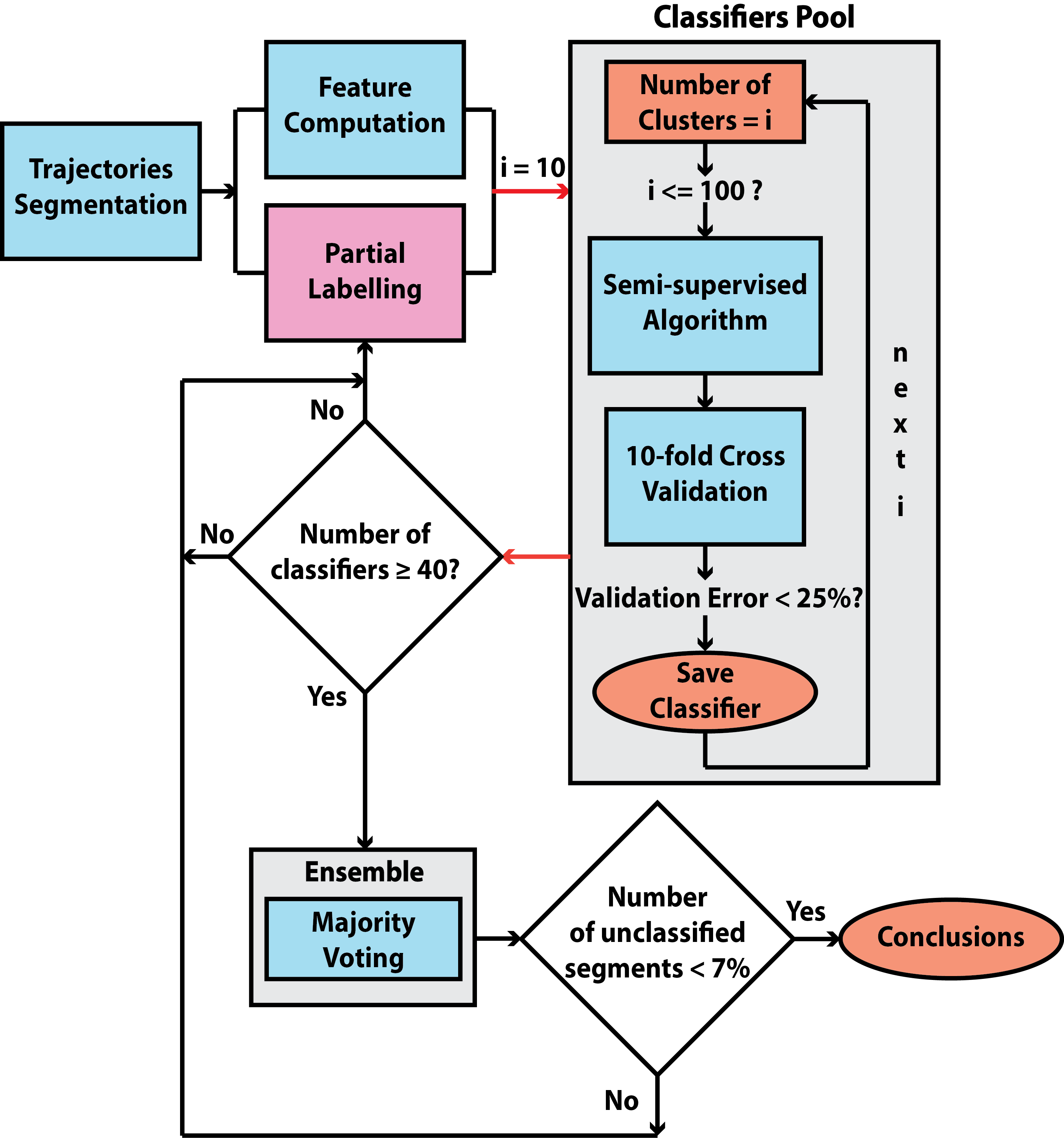}
	\end{center}
	\caption{\textbf{Workflow diagram illustrating the analysis procedure.} Cyan boxes indicate automatic process; orange boxes indicate objects of importance or results; Partial Labelling box (magenta) implies extensive user interaction with the process; grey boxes group the processes taking part in the classification procedure. After the trajectories segmentation, eight trajectories features for each segment are computed and a certain number of segments is manually labelled. Afterwards a pool of classifiers is generated. `Strong' classifiers (cross validation error $<25\%$) are then selected from the pool and work together (majority voting) as a team (ensemble) to produce the classification results. Throughout the process the labelling quality is constantly assessed and in case of weak classification results we go back to the labelling stage.} \label{fig:overview}
\end{figure}

In this work, instead of searching for an optimal number of clusters and attempting to generate an optimal classifier, we select to generate a pool of `strong' classifiers whose `goodness' is assessed based on the 10-fold cross validation error. The strong classifiers generated in that way are then used to form an `ensemble' which uses majority voting to reach a classification decision. The two conditions of having both strong and diverse classifiers are essential in majority voting in order to reach an optimal classification solution \cite{sharkey1997diversity,zhou2002ensembling}. This will be discussed in more detail later. In order to assess the labelling procedure (if enough and consistent labels have been provided) the criterion of having a minimum of 40 strong classifiers has been added prior to majority voting. Finally, the classification result of the ensemble is expected to have a low percentage of unclassified segments (less than 3\%) because, since the classifiers are diverse, they will do different errors or will fail to classify different segments. Thus if they work together and form an ensemble, the individual errors will be compensated by the correct responses of the other members of the ensemble \cite{sharkey1997diversity}. A diagram of the procedure is illustrated in figure \ref{fig:overview}.

\subsection{Trajectories Segmentation and Partial Labelling}
To assign one trajectory to multiple classes, we earlier proposed the division of the full animal swimming paths into segments \cite{gehring2015detailed}. In our method each segment overlaps significantly with its previous one (percentages of 70\% and 90\% have been performed on this analysis) to make sure that important information is not lost due to an unfavourable segmentation. The segment length was empirically selected to be equal to or slightly longer than one arena diameter. If the segment length is too short it might be difficult to identify to which class segments belong; if it's too long it might happen that more than one class of behaviour is represented. The latter case can be seen in our results, where the large segment length (3 times the arena radius) causes some classes to be overshadowed by the more common classes (refer to figure \ref{fig:res_main_1}).

In this study, nine predefined strategies were adopted (see Classes of Behaviour). We have found empirically that the amount of data that needs to be labelled should be roughly between $8\%$ to $12\%$ of the total segment number but the exact value depends greatly on the dataset under investigation. As a rule of thumb, if fewer labels are provided then the classification results will be poor in the sense that a lot of segments will remain unclassified or fall under the wrong class. Since the labelling procedure is prone to error and subjectivity a number of validation criteria have been implemented throughout our analysis (see figure \ref{fig:overview}).

\subsection{Classification Boosting with Majority Voting}
The classification boosting is an ensemble technique that is based on the idea that many weak learners can be converted to a strong learner \cite{kearns1993cryptographic}. In machine learning terms an ensemble of weak classifiers (classifiers that make mistakes) can be used to form a strong classifier (classifier that makes far less mistakes) by combining each individual's opinion \cite{gerecke2003common,jurek2011classification}. This approach has been used in various classification tasks (see Oza et al. \cite{oza2008classifier} for a survey) and in addressing complex real-world problems, when single algorithmic classification solutions are unable to achieve high performances \cite{acharya2011c}. 

One way to perform classification boosting is through majority voting \cite{gerecke2003common}: many classifiers form an ensemble, vote for the class of each datapoint and the class with the more votes wins. The output of the ensemble is expected to have an improved accuracy since individual errors of each classifier are compensated by the correct responses of the other members of the ensemble \cite{sharkey1997diversity}. However, in order to achieve such outcome, the classifiers need to be diverse (they should not share the same errors) \cite{schapire1990strength,gerecke2003common}. In fact, according to \cite{sharkey1997diversity} and \cite{zhu2015use}, diversity is not enough to ensure that randomly selected classifiers will achieve high classification accuracy, under the scenario that all the classifiers are arbitrary week. Classifiers have to also be strong meaning that they should be sufficiently accurate on their own (\cite{zhu2015use}) and \cite{ruta2002theoretical} indicate an accuracy of at least 50\%).


\subsubsection{Majority Voting Implementation}
In our framework, we need to classify different trajectory segments into animal behavioural classes (strategies) having only a partial set of labelled data. The classification is parametrized by the target number of clusters of the clustering algorithm, a value that is difficult to estimate a priori. In order to overcome this problem we therefore generate a number of classifiers by providing different numbers of target clusters in succession; in the end of this process a pool of classifiers is generated. We use the 10-fold cross validation \cite{varma2006bias} process to evaluate different number of target clusters (10 to 100). Only classifiers with a validation error lower than 25\% are used to form an ensemble (for more information about the 10-fold cross validation procedure refer to the appendix). We set the minimum amount of required classifiers that fulfill this criteria to 40. The reasoning behind this process is that we require a sufficient number of `strong' classifiers. For the majority voting, we consider the simple scheme where the vote of each classifier has the same weight \cite{liaw2002classification,bouziane2011profiles} and that in case of a draw the datapoint (segment) is marked as unclassified.

\subsection{Framework Validation}
\begin{figure}[h!]
	\begin{center}
		\includegraphics[width=0.6\textwidth]{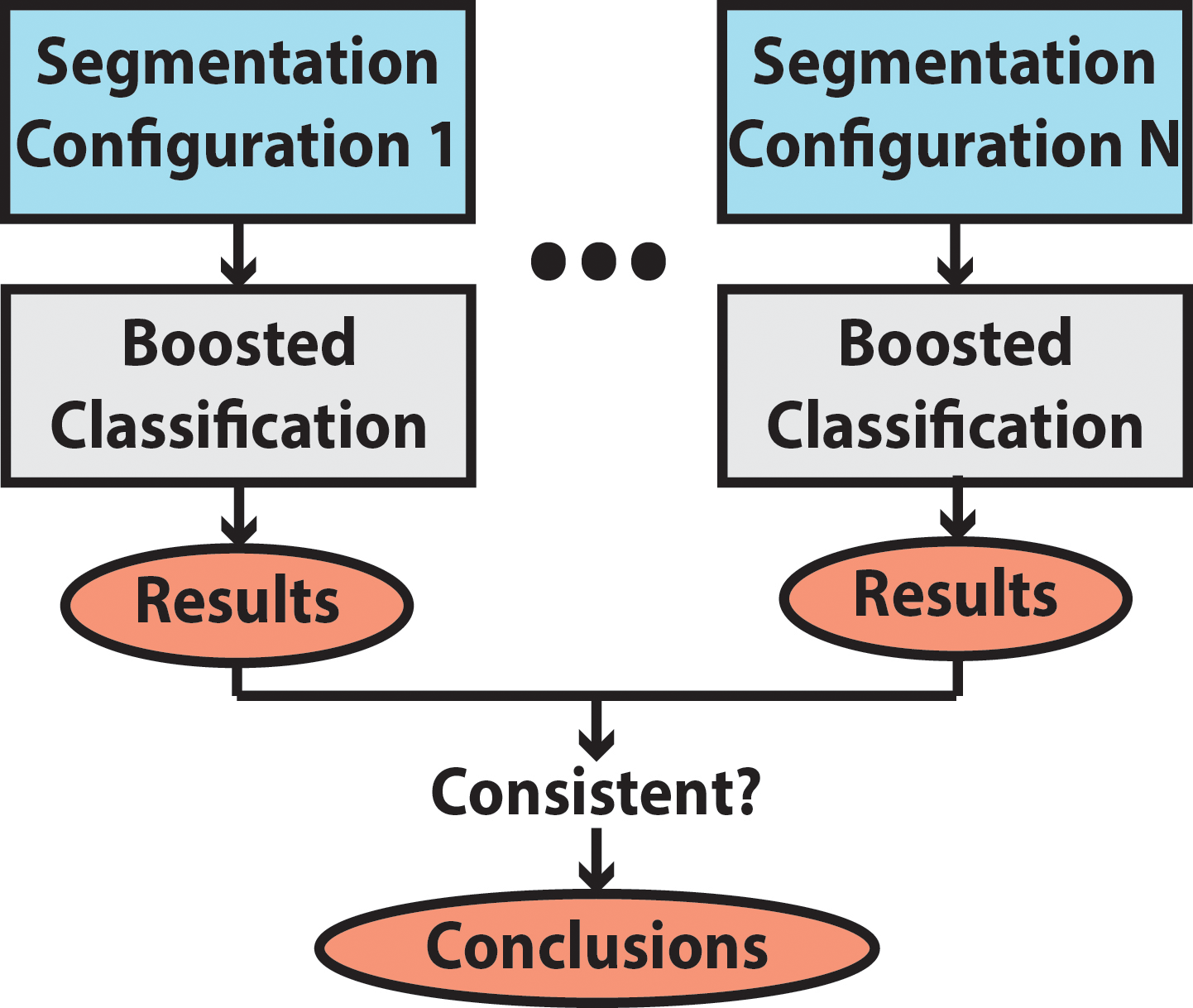}
	\end{center}
	\caption{\textbf{Validation stage.} Cyan boxes indicate automatic process; orange boxes indicate objects of importance or results; grey boxes indicate the classification procedure. Different segmentation configurations (varied segments length and overlap) have been tested in order to prove that the conclusions are consistent (they are not based on a particular segmentation) and to define the bounds for the segment length between which we have the same analysis conclusions. Refer to table \ref{table:summary1} for the properties of each configuration.}\label{fig:validation1}
\end{figure}

The validation process of the framework (see figure \ref{fig:validation1}) aims to prove that different segmentation configurations are able to lead to the same conclusion (segmentation robustness). More specifically, it was used to define the bounds (error margins) for the segment length between which we have consistent analysis conclusions. 

Throughout the validation stage and the data analysis process as a whole we considered certain statistics and validity measurements which will be described next.

\subsection{Statistics}

The non-parametric Friedman test \cite{siegel1956nonparametric} was used for the analysis of variance of each strategy between the two animal groups. This test was selected because the data are not normally distributed and because of its ability to control the variability among subjects over the different observations \cite{theodorsson1987friedman}. 

For our analysis the null hypothesis is that there in no difference between the two animal groups (stress and control) over each one of the strategies (refer to section \ref{classes_of_beh}) as well as  over the number of times that the animals change their behaviour within single trials (strategy transitions). Small p-values ($<0.05\%$) generated by the Friedman test lead us to discard the null hypothesis that the results are identical and that any differences are only due to chance (random sampling).

In addition to the Friedman test, the 95\% confidence intervals of a binomial distribution \cite{wallis2013binomial} are being used, where the significance of a specific classification, as judged by each of the classifiers that form the ensemble, is viewed as a random process generating one (significant differences) or zero (non-significant differences). In more detail, the confidence intervals indicate our confidence that the classifiers forming the ensemble are on average pointing to the same conclusion as the ensemble (i.e. the majority agrees that there is significant difference over strategies or strategies transitions). Given that the Friedman test can have two outcomes, we hypothesise the outcomes to be the result of a binomial distribution. We require that the 95\% confidence intervals to be clearly above 0.5 (or 50\%) in order to be confident that the result in not due to chance \cite{brown2001interval,peck2012statistics}. 

\subsection{Validity Measures}
In the section below we discuss the classifier diversity and the percentage of unclassified segments which are our key criteria use to validate the classification process. 
 
\subsubsection{Classifier Diversity} 
To evaluate the diversity of the classifiers, we assess the percentage of their agreement for the class of each segment. The result is a symmetric matrix with rows and columns representing the classifiers and each element shows the percentage of segments for which two classifiers agree on the assigned class. The diagonal values of this matrix equal to 100 as each classifier is in 100\% agreement with itself (refer to figure \ref{fig:agreement_matrix} in the appendix section for an example of an agreement matrix). An overall agreement can be computed by averaging the upper or lower triangular of the matrix. In addition, we consider the average cross validation error (accuracy) over the classifiers. In order for the classifiers to be both diverse and strong it is expected that they should have an average percentage of agreement well below 100\% (in our case around 60\%) and low cross validation error (refer to Table \ref{table:summary1a}).


As it has been previously reported \cite{sharkey1997diversity}, ensembles have far less variance in comparison with the classifiers thus it is expected to have much higher agreement. To demonstrate this observation, we generated a number of ensembles by picking classifiers at random from the pool. Afterwards we performed the same statistical measurement of agreement for the ensembles, similar to the one described for the classifiers. In contrast to the classifiers, the ensembles have high agreements among them (more than 80\%) and nearly nullify the cross validation error of the classifiers \ref{table:summary1a}. However, since in our method the cross validation was used for both tuning and testing \cite{gehring2015detailed}, additionally we manually assess the error of the ensembles on two out of the four segmentations (see the appendix section).

\subsubsection{Percentage of unclassified segments}
A useful measure for the quality of the classification is the percentage of unclassified segments. For certain segments, it is expected that none of the classifiers in the ensamble will be able to determine a class, or that there could be a draw for segments that transit between classes (refer to the results section Table \ref{table:segments_count}). This, however, does not have an impact on the results consistency (see figure \ref{fig:res_main_1}). 

\subsection{Mapping Segment Classes to the Full Swimming Paths}
The classification has been performed on overlapping segments of the animals' swimming paths, and therefore we need to map them back to the whole trajectories. 

As a first approach, we considered the classified segments as continuous parts of the trajectories ignoring the overlap percentage. This method provides consistent results on the significant differences of the strategies but fails to detect differences on strategies transition between groups (refer to figure \ref{fig:no-smoothing-class} in the appendix section). The reason for this is that sparse segments within each swimming path fall under different classes thus viewing them as a sequence leads to an overestimation of transitions (a transition happens when a segment falls under a different class after a sequence of segments that fall under the same class).

To address this limitation, we use a smoothing technique with parameters independent of the segmentation choices. This was done for the following reason: (i) to avoid subjective conclusions based on a specific segmentation configuration and (ii) to be able to directly compare different segmentations. In more detail, given that $R$ equals to the radius of the arena, the swimming paths are now divided into intervals of length $R$. Each of the intervals is assigned to a certain class based on a weighed voting of all the overlapping segments. The mathematical expression for this operation is shown in equation \ref{eq:smoothing},

\begin{equation}\label{eq:smoothing}
C_{T_i} \equiv arg_{c_k}max\sum_{\binom{S_j \in c_k}{T_i\cap S_j\neq\varnothing}} w_k \cdot e^{-\frac{d_{ij}^2}{2 \cdot \sigma^2}} 
\end{equation}

\noindent where $T_i$ is the $i_{th}$ interval, $d_{i,j}$ is the distance from the centre of the $j_{th}$ segment ($S_{j}$) overlapping with the $i_{th}$ interval to the centre of the $i_{th}$ interval, $c_k$ is the $k_{th}$ segment class and $w_k$ is a class weight normalised so that $\sum{w_k}=1$. The sum is to be taken over the segments intersecting with the interval $T_i$, belong to class $c_k$ (unclassified segments are excluded) and fulfill the threshold requirement $e^{-\frac{d_{ij}^2}{2 \cdot \sigma^2}} >= 0.14$, where $\sigma$ is the variance of the Gaussian and the value 0.14 is obtained when $d_{ij} = 2 \cdot \sigma$. The reason for the latest requirement is to create a cutoff for the segments that are too far away from the centre of the interval. The parameter $\sigma$ controls the weight of the vote of each segment based on its distance from the interval and in our analysis it was set equal to $R$ in order to achieve proportionality with the arena dimensions (other values have also been tested, refer to the appendix section). Finally, the class weight $w_k$ was defined as $w_k = \frac{1}{P(c_k)}$, where $P(c_k)$ is the percentage of segments belonging to class $k$. The intuition for setting the class weights inversely proportional to the amount of segments that fall under each class was to prevent more rare classes to be overshadowed by more common ones. To prevent having too small or too large class weights the bounds of [0.01 0.5] were set, which means that if less than 1\% or more than 50\% of the segments fall under a certain category then this class will receive weight equal to 0.5 or 0.01 respectively.

\subsection{The RODA Software}
\begin{figure}[h!]
	\begin{center}
		\includegraphics[width=\linewidth]{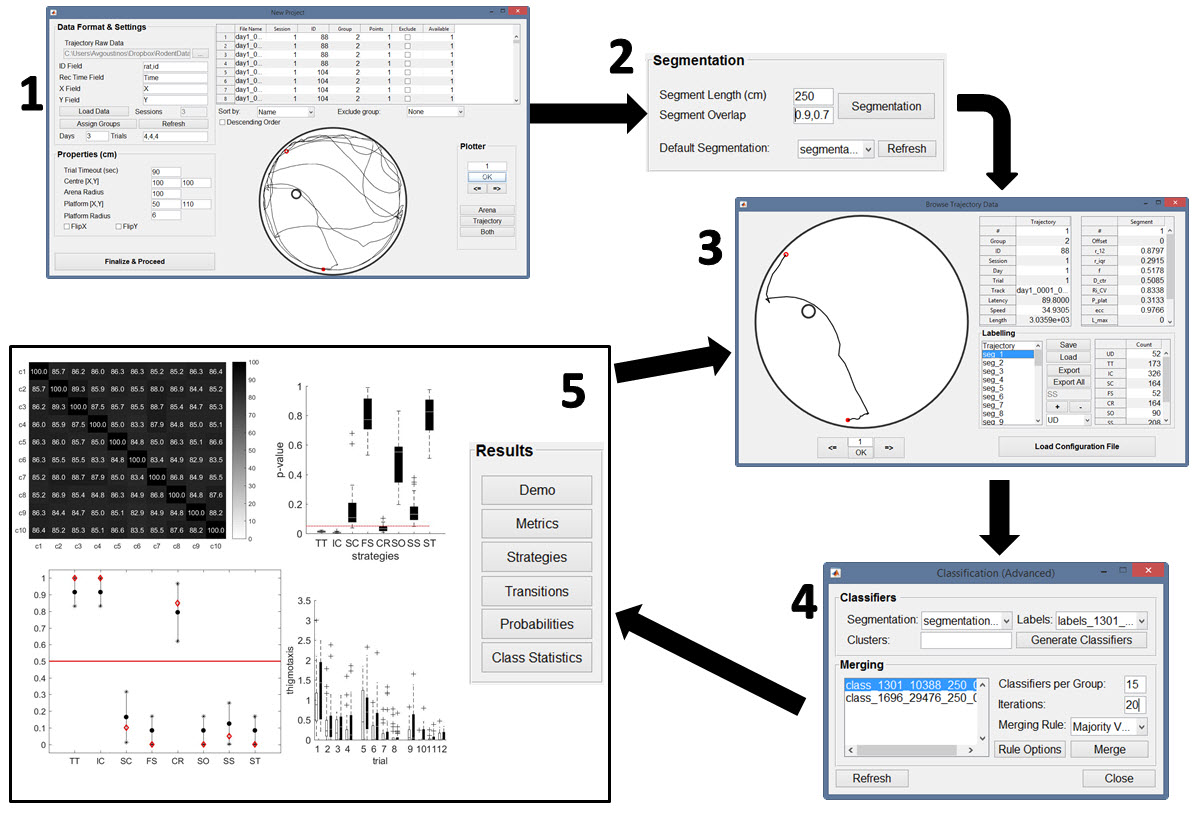}
	\end{center}
	\caption{\textbf{Screenshots of the software.} Each window is numbered to denote a separate stage of the workflow, which consists of: \textbf{(1)} the data input GUI, which is used to load the trajectory data extracted from Ethovision and select the specific tracks that will be used in the analysis; \textbf{(2)} the segmentation panel, which offers fully control over the segmentation options; \textbf{(3)} the labelling GUI, which offers visualisation of the whole trajectories and their segments allowing easy labelling of the segments; \textbf{(4)} the classification GUI, which contains options to tune various parts of the classification process; \textbf{(5)} the results panel; which generates the analysis results. The results are generated in both graphical as well as text format. The user has also control over the output format of the image files as well as the elements of the generated figures such as text size, line width, etc. The arrow connecting \textbf{(5)} with \textbf{(3)} indicates that if the analysis results are not consistent then we need to go back to the labelling stage and provide more or better labels.}\label{fig:software}
\end{figure}

The RODA software \cite{avgoustinos_vouros_2017_1117837} (shown in Figure \ref{fig:software}), which implements this framework, consists of a series of graphical user interfaces (GUIs) which offer straightforward analysis of trajectory data extracted from the Noldus Ethovision System \cite{noldus2001ethovision}. Every stage of the process can be tuned to meet the user's need. The generated figures can be exported into a variety of different image formats (JPEG, TIFF, etc.) while the numerical data depicted in the figures are also saved in Comma Separated Values (CSV) file format in case the user wishes to generate the figures using a different software (e.g. Microsoft Excel).

The software is entirely written on MATLAB and uses a modified version of the WEKA library \cite{frank2016weka} written in JAVA which is known as WekaUT (for more information refer to \href{http://www.cs.utexas.edu/users/ml/risc/code/}{http://www.cs.utexas.edu/users/ml/risc/code/}) for the clustering procedure.

\subsection{Morris Water Maze Experiment and Data Properties} 
The data have been collected from experiments performed at the Laboratory of Behavioural Genetics, EPFL at Lausanne, Switzerland. All procedures were conducted in conformity with the Swiss National Institutional Guidelines on Animal Experimentation and approved by a license from the Swiss Cantonal Veterinary Office Committee for Animal Experimentation.

The water maze had a diameter of $200cm$ with a submerged platform of diameter $12cm$. The recordings of the animals trajectories were performed by using the tracking software, Noldues EthoVision \cite{noldus2001ethovision} version 3.1. The dataset contains 57 rats, 30 of which were inducted into stress at peripubertal age \cite{marquez2013peripuberty} and 27 of which were the control group. A total of 12 trials were performed per animal divided into 3 consecutive days with 4 trials per day. The timeout of each trial was 90 seconds and if the animal failed to find the platform within the time limit it was guided to it. The inter-trial interval between the trials of the same day was only a few minutes. The starting position of the animals was altered between trials.

\subsection{Classes of Behaviour}\label{classes_of_beh}
The choice of the classes of behaviours (strategies) in our analysis is motivated by previous studies (e.g. \cite{wolfer1992new,wolfer1998spatial,graziano2003automatic}) which have observed and reported stereotypical animal behaviours inside the MWM (for an example of each strategy refer to the appendix section). 

\noindent\textbf{Thigmotaxis.} The animal moves exclusively on the periphery of the arena and most of the time it touches the walls of the arena.

\noindent\textbf{Incursion.} The animal starts to distant itself from the arena periphery with visible inward movements. 

\noindent\textbf{Scanning.} A behaviour associated with random searches focused in the centre of the pool. Another characteristic of this behaviour is that the animal rapidly turns away from the arena walls if it touches them \cite{graziano2003automatic}. 

\noindent\textbf{Focused Search.} This behaviour is also associated with random searches but here the animal actively searches a particular small region of the arena.

\noindent\textbf{Chaining Response.} A behaviour first observed in the study of Wolfer et al.  \cite{wolfer2000dissecting} where the animal appears to have memorised the distance to the platform from the arena wall and swims circularly in order to find it.

\noindent\textbf{Self Orienting.} The animal performs a loop and orients itself inside the arena \cite{graziano2003automatic}.

\noindent\textbf{Scanning Surroundings.} The animal crosses a region very close to the platform of the arena but moves away \cite{gehring2015detailed}.

\noindent\textbf{Scanning Target.} The animal actively searches for the arena by swapping paths around it.

\noindent\textbf{Direct Finding.} The animal navigates straight to the platform.

\subsection{Abbreviations}

In some figures the following strategies abbreviations are being used: Thigmotaxis (TT), Incursion (IC), Scanning (SC), Focused Search (FS), Chaining Response (CR), Self Orienting (SO), Scanning Surroundings (SS), Target Scanning (ST), Direct Finding (DF). In addition, we have analysed the number of times that the animals change their behaviour within single trials (Strategy Transitions or tr).

\section{Results}

\subsection{Advantages of Trajectory Segmentation Analysis}

Our methodology finds quantitative behavioural differences in comparison with standard metrics on the full swimming paths of the animals. It is able to detect additional significant differences between the strategies employed by the two animal groups in comparison to the categorisation of the whole animals trajectories.

\begin{figure}[h!]
	\begin{center}
		\includegraphics[width=\linewidth]{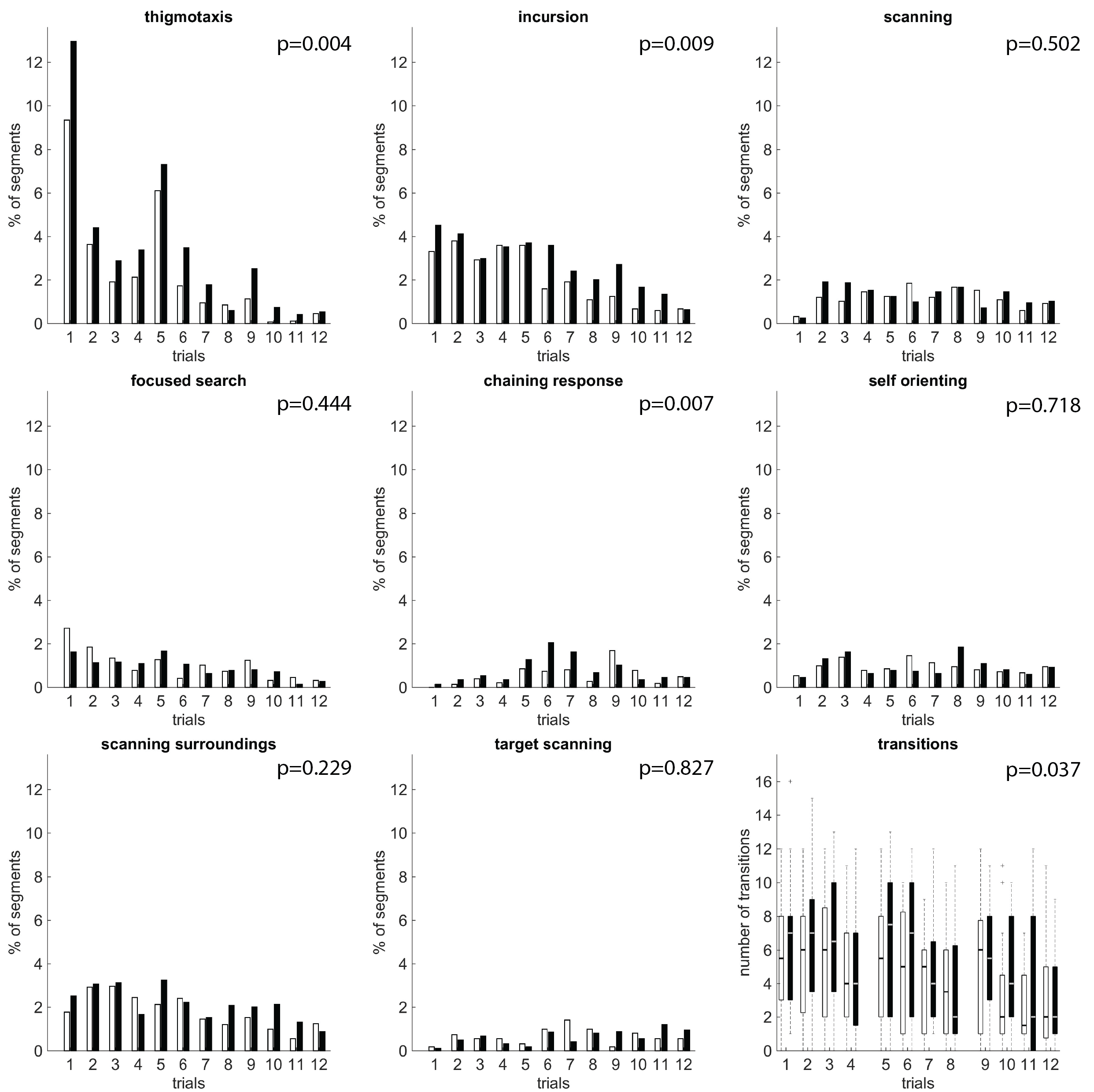}
	\end{center}
	\caption{\textbf{Percentage of segments falling under each strategy for the stress (black) and control (white) animal groups over each trial.} All the animals were tested for a set of 12 trials divided in 3 sessions (days). Each segment (now path interval) is considered to be of length equal to the length of the arena radius (100cm). For the transitions: bars represent the first and third quartiles of the data; the black (control group) or white (stress group) horizontal lines is the median, crosses are the outliers and whiskers indicate the minimum and the maximum values. These results were generated by using segmentation length of 250cm and 90\% overlap; for the classification an ensemble of classifiers was created by using classifiers with validation error less than 25\%; equivalent results of the other three segmentations can be found in the appendix section. The Friedman test p-value (shown on the top right) was used to compare both animal groups for the complete set of trials. According to the plots Thigmotaxis and Incursion strategies show a clear difference in favour of the stress groups along with Chaining Response. The number of transitions between strategies shows that the stress animals change their behaviour more ofter than control animals within single trials.}\label{fig:res_main_2}
\end{figure}

In more detail, we report that the two animal groups (stress and control) differ on the strategies of Thigmotaxis, Incursion and Chaining Response and strategies transitions (see figure \ref{fig:res_main_1}) in favour of the stress group meaning that stressed animals implement these strategies and transit between different strategies more often than the control animals (see figure \ref{fig:res_main_2}).

\begin{figure}[h!]
	\begin{center}
		\includegraphics[width=\linewidth]{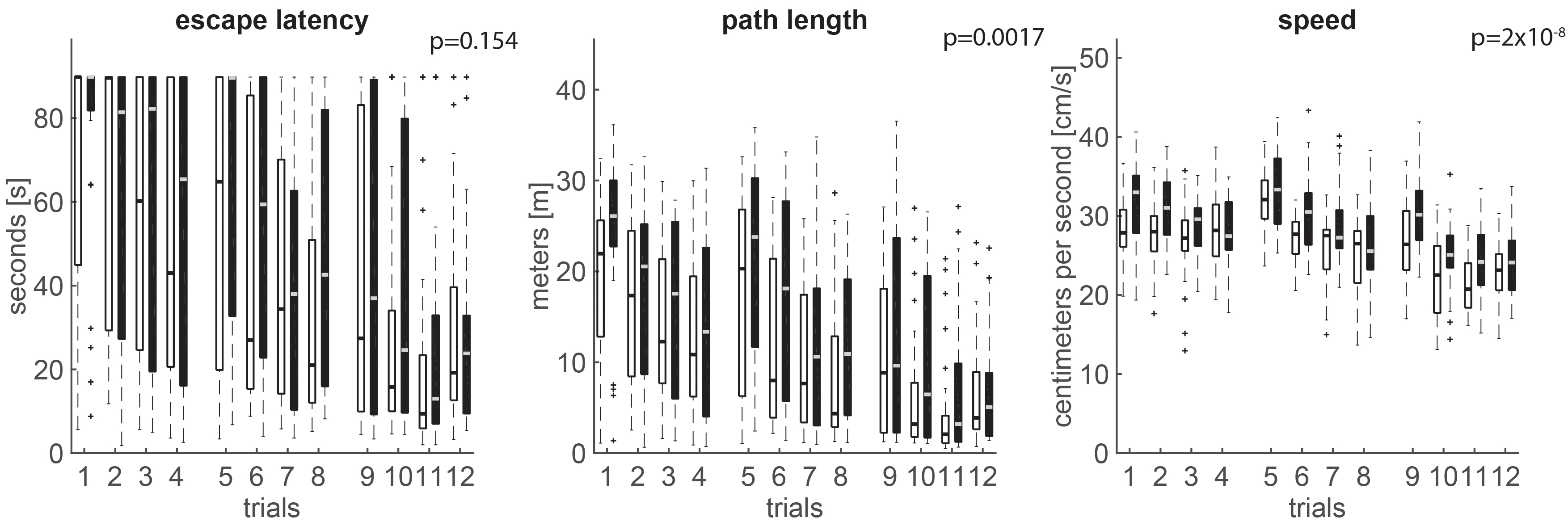}
	\end{center}
	\caption{\textbf{Full swimming path standard metrics for the stress (black) and control (white) animal groups.} All the animals were tested for a set of 12 trials divided in 3 sessions (days). Bars represent the first and third quartiles of the data; the grey line that splits the bars represents the median, crosses are the outliers and whiskers indicate the minimum and the maximum values. The Friedman test p-value over the trials is shown on the top right of each plot. Stress animals find the platform as fast as the control group (\textbf{A. escape latency}, p-value: $0.15$) even though they run faster (\textbf{B. speed}, p-value: $2$x$10^{-8}$) and sweep (on average) longer swimming paths (\textbf{C. path length}, p-value: $0.0017$) within the trials than the control group.}\label{fig:metrics}
\end{figure}

\begin{figure}[h!]
	\begin{center}
		\includegraphics[width=\linewidth]{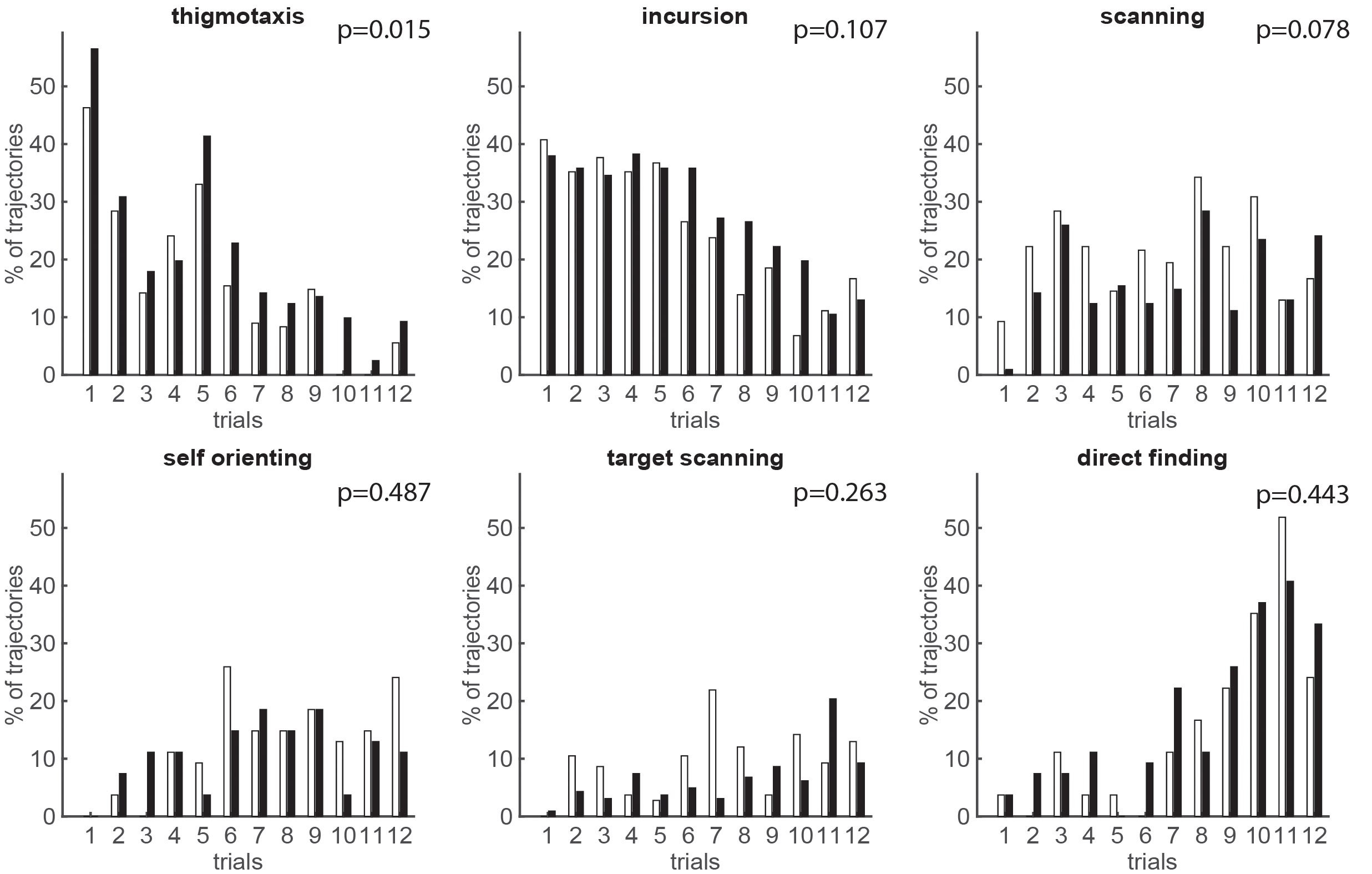}
	\end{center}
	\caption{\textbf{Manual classification of the full swimming paths.} In the manual classification of the full swimming paths significant differences in certain categories (Focused Search, Chaining Response and Scanning Surroundings) couldn't be identified. Significant difference (p-value equal to 0.015) was detected only for the Thigmotaxis strategy. White bars: control group; Black bars: stress group. The two groups were compared over the complete set of trials using the Friedman test (shown on the top right corner of each graph.)}\label{fig:full_swimming_paths}
\end{figure}

Commonly used measurements of learning (animal speed, escape latency and path length) suggest that there is a significant difference among the two animal groups in the sense that the stress animals are faster and swap longer paths within the trials but they still fail to find the platform in less time than the control animals (see figure \ref{fig:metrics}). Our analysis suggests that the reason for this phenomenon is that stressed animals tend to use more low level strategies (Thigmotaxis and Incursion) which lower their chances on finding the platform since they spent most of the time close to the arena periphery. More high level cognitive strategies such as scanning target do not show any significant difference between the two groups except the Chaining Response strategy. The latter implies that stress animals haven't memorised exactly the location of the platform but its distance to the wall; so they swim at that distance in hope to find it by chance. Stressed animals also change strategy more often than non-stressed animals.  These results are relevant to studies such as \cite{aston2000locus,luksys2009stress,luksys2011neural} which suggests that high levels of stress lead to weak attention and frequent behavioural switches. Categorisation of the full animal swimming paths failed to detect distinctive strategies such as Chaining Response; it only manages to capture differences for the Thigmotaxis strategy (see figure \ref{fig:full_swimming_paths}).

\cleardoublepage


\subsection{Robustness across different segmentations}

Four different segmentations (with different segment length and/or segment overlap) were performed (refer to table \ref{table:summary1}). For each segmentation, we generate a pool of classifiers which consisted of classifiers with cross validation error lower than 25\%. These classifiers were used to form an ensemble and the final classification result was computed based on majority voting within the ensemble (refer to Table \ref{table:summary1} for concise information about each segmentation configuration and Table \ref{table:summary1a} for measurements about the classifiers diversity and the majority voting benefits). 

\begin{figure}[h!]
	\begin{center}
		\includegraphics[width=\linewidth]{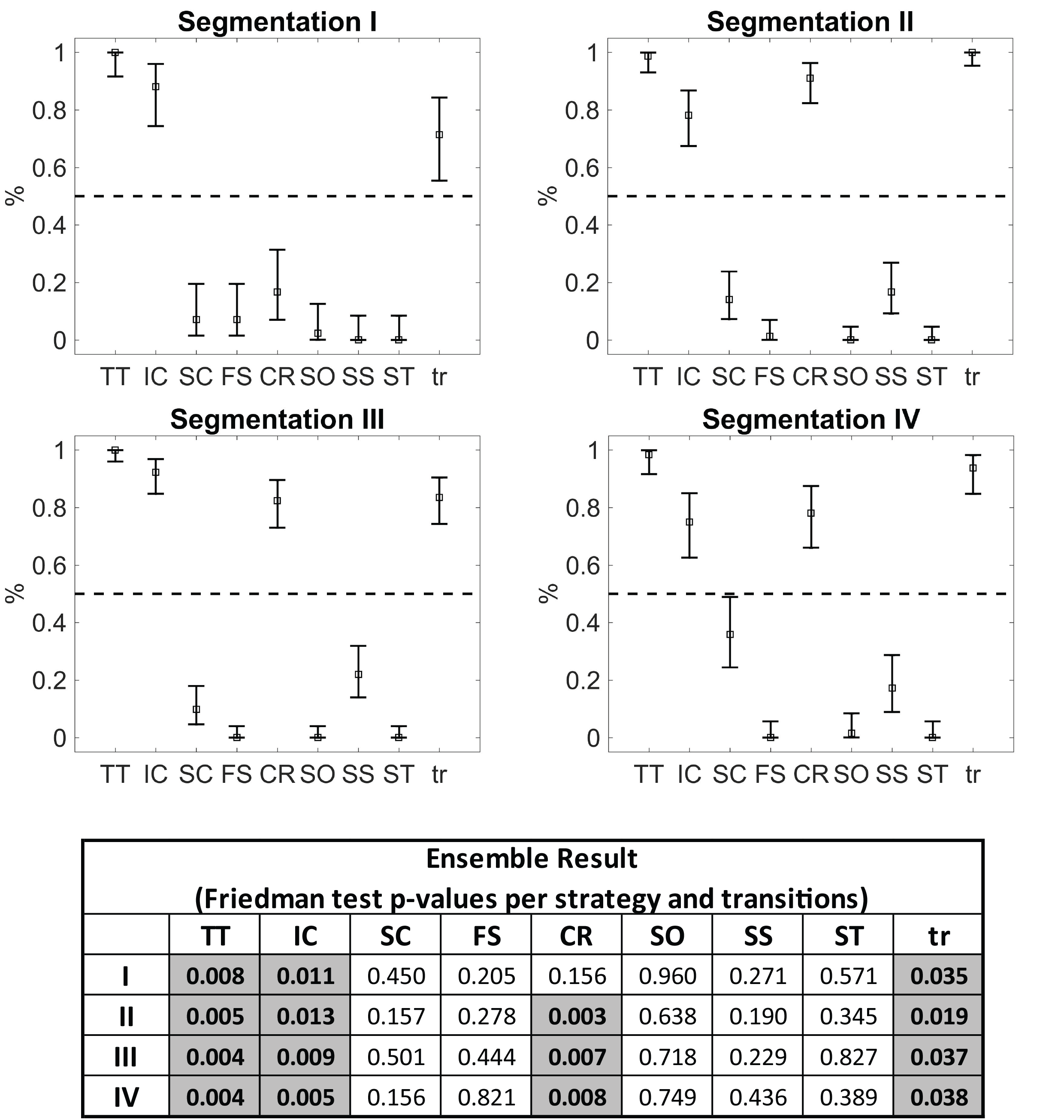}
	\end{center}
	\caption{\textbf{Conclusive results from the classification of each segmentation (see Table \ref{table:summary1}).} Each plot shows the 95\% binomial confidence intervals for the classifiers of each segmentation regarding their agreement on the significant difference between the two animal groups for each strategy and the strategy transitions(Friedman test p-value $< 0.05$). Squares indicate the mean of the classifiers; errorbars represent the 95\% confidence intervals; the dashed line indicates the threshold of interest (0.5 or 50\%). Confidence intervals clearly above 0.5 (or 50\%) confirm that there is indeed a significant difference between the the two animal groups on the strategies and the strategy transitions. We see that in three cases (Segmentations II, III, IV) the two animal groups show significant differences in the strategies of Thigmotaxis, Incursion and Chaining Response and transition between strategies. Segmentation I failed to capture the significant difference on the Chaining Response because of the lengthy segments which caused this strategy to be overshadowed by other strategies and disappear (refer also to table \ref{table:segments_count}). The table below the plots shows the Friedman test p-values for the classification result of the ensembles. Segmentation configurations are arranged in columns and strategies in rows; each element has the relevant p-value and grey cells indicate significant difference (p-value $< 0.05$).}\label{fig:res_main_1}
\end{figure}

Three out of four segmentation configurations (with segment lengths 250cm and 200cm) led to the conclusion that the two animal groups (stress and control) have significant difference on the strategies of Thigmotaxis, Incursion and Chaining Response and strategies transitions (see figure \ref{fig:res_main_1}) in favour of the stress group meaning that stress animals implement these strategies and transit between different strategies more often than the control animals (see figure \ref{fig:res_main_2}). One out of four segmentations (segment length 300cm) fail to capture significant difference in the Chaining Response strategy and a probable reason is that the segment length is too large (3 times the arena radius), thus strategies that are more rare and significantly smaller are overshadowed by more common ones (e.g., Chaining Response may be overshadowed by Scanning Surrounding or Thigmotaxis, refer to \ref{table:segments_count}). This is an issue introduced already during the labelling procedure. For the Segmentation 1 only 0.67\% of the samples were single-labelled as {\it chaining response} vs 1.58\%, 0.72 \% 1.06 \% for the Segmentations 2 to 4 correspondingly. The larger segment makes it more difficult for the human expert to distinguish rare classes that are adjoint to frequent ones.
\begin{table}[h!]
	\centering
	\begin{tabular}{|l|c|c|c|c|}
		\hline
		& \begin{tabular}[c]{@{}c@{}}Segmentation\\ I\end{tabular} & \begin{tabular}[c]{@{}c@{}}Segmentation\\ II\end{tabular} & \begin{tabular}[c]{@{}c@{}}Segmentation\\ III\end{tabular} & \begin{tabular}[c]{@{}c@{}}Segmentation\\ IV\end{tabular} \\ \hline
		\begin{tabular}[c]{@{}l@{}}Segment \\ Length\end{tabular}       & 300                                                      & 250                                                       & 250                                                        & 200                                                       \\ \hline
		\begin{tabular}[c]{@{}l@{}}Segment \\ Overlap\end{tabular}      & 70\%                                                     & 70\%                                                      & 90\%                                                       & 70\%                                                      \\ \hline
		\begin{tabular}[c]{@{}l@{}}Number of\\ Segments\end{tabular}    & 8847                                                     & 10388                                                     & 29476                                                      & 13283                                                     \\ \hline
		\begin{tabular}[c]{@{}l@{}}Number of\\ Segments Labelled\end{tabular}      & \begin{tabular}[c]{@{}c@{}}988\\ (12\%)\end{tabular}   & \begin{tabular}[c]{@{}c@{}}1261\\ (12\%)\end{tabular}   & \begin{tabular}[c]{@{}c@{}}2445\\ (8\%)\end{tabular}     & \begin{tabular}[c]{@{}c@{}}1227\\ (9\%)\end{tabular}    \\ \hline
		\begin{tabular}[c]{@{}l@{}} Total number\\ of labels\end{tabular}    & 1022                                                     & 1313                                                     & 2568                                                      & 1232                                                     \\ \hline		
		\begin{tabular}[c]{@{}l@{}}Number of\\ Classifiers\end{tabular} & 42                                                       & 78                                                        & 91                                                         & 64                                                        \\ \hline
	\end{tabular}
	\caption{\textbf{Parameters for the classification of four different segmentation configurations with variable segment lengths and overlaps.} For each segmentation, a pool of classifiers was generated with varied number of classifiers depending on the cross-validation error (only classifiers with error lower than 25\% were present). The percentage of the manually labelled segments was between 8\% and 12\%. Multiple labels could be given to each segment; in this study no more than two labels were given simultaneously to a segment.}\label{table:summary1}
\end{table}

In order to validate the results for all four segmentations (see Table \ref{table:summary1}) we chose to form 21 ensembles each one of which were consisted of 11 classifiers. This also allowed us to compare the agreement among classifiers and among ensembles (see Table \ref{table:summary1a} ensemble average agreement). The classifiers used to form each ensemble were picked at random each time. 

We compared  the performance of the classifiers, the ensemble and the multiple ensembles formed by random sample of classifiers. Table  \ref{table:summary1a} shows the relevant results on the last stage of analysis, after the smoothing function has been applied and the segments had been mapped to the full swimming paths; this detail is important because the smoothing procedure increases the performance of classifiers (for the statistical analysis prior to the smoothing function refer to Table \ref{table:presummary1a} in the appendix section). As expected, ensembles have higher accuracy, lower percentage of unclassified segments and higher percentage of agreement among them in comparison to the classifiers. However, since in our method the cross validation was used for both tuning and testing, additionally we manually assess the error of the ensembles on two out of the four segmentations (see the appendix section).

\begin{table}[h!]
	\centering
	\begin{tabular}{|l|c|c|c|c|}
		\hline
		\multicolumn{1}{|c|}{}                                               & \begin{tabular}[c]{@{}c@{}}Segmentation\\ I\end{tabular}            & \begin{tabular}[c]{@{}c@{}}Segmentation\\ II\end{tabular}           & \begin{tabular}[c]{@{}c@{}}Segmentation\\ III\end{tabular}          & \begin{tabular}[c]{@{}c@{}}Segmentation\\ IV\end{tabular}           \\ \hline
		& \multicolumn{4}{c|}{Classifiers}                                                                                                                                                                                                                                                      \\ \hline
		\begin{tabular}[c]{@{}l@{}}Error (\%) \\ {[}min-max{]}\end{tabular}  & \begin{tabular}[c]{@{}c@{}}16.8\\ {[}5.4 24.9{]}\end{tabular}   & \begin{tabular}[c]{@{}c@{}}17.5\\ {[}3.7 25.0{]}\end{tabular}   & \begin{tabular}[c]{@{}c@{}}13.9\\ {[}1.8 21.5{]}\end{tabular}   & \begin{tabular}[c]{@{}c@{}}18.0\\ {[}7.3 24.9{]}\end{tabular}   \\ \hline
		\begin{tabular}[c]{@{}l@{}}Unclassified (\%)\\ Segments\end{tabular} & 2.5                                                              & 2.5                                                               & 1.3                                                               & 3.7                                                             \\ \hline
		Agreement (\%)                                                       & 58.7                                                              & 61.0                                                              & 59.6                                                             & 56.3                                                            \\ \hline
		& \multicolumn{4}{c|}{Ensemble(s)}                                                                                                                                                                                                                                                      \\ \hline
		\begin{tabular}[c]{@{}l@{}}Error (\%)\end{tabular}                    & 0.0                                                    & 0.2                                                     & 0.0                                                      & 0.0                                                     \\ \hline	
		\begin{tabular}[c]{@{}l@{}}Unclassified (\%)\\ Segments\end{tabular} & 0.0                                                              & 0.0                                                             & 0.0                                                              & 0.1                                                              \\ \hline
		Agreement (\%)                                                       & 83.4                                                             & 82.6                                                              & 82.3                                                              & 80.0                                                              \\ \hline
	\end{tabular}
	\caption{\textbf{Classification statistics (average) for the four segmentation configurations of Table \ref{table:summary1} and benefits of majority voting.} \textbf{(1)} Error: the 10-fold cross validation was used in order to select `strong' classifiers based on their validation error.  10-fold cross validation was also used to compute the average accuracy of the `strong' classifiers and the accuracy of the ensemble (in case of the ensemble, the same folds used by the classifiers were re-used). The ensemble significantly benefits the classification accuracy. Since in our method the cross validation was used for both tuning and testing we manually assess the error of the ensembles on two out of the four segmentations (see the appendix section). \textbf{(2)} The percentage of unclassified segments was computed separately; since the classifiers are `strong' only a few segments remain unclassified, nevertheless the ensemble almost totally nullifies the unclassified segments. \textbf{(3)} The average agreement between the classifiers was computed by first calculating the percentage of agreement within each pair (we have agreement when two classifiers have assigned the same label on a particular segment) and then averaging all the agreements together (refer to Validity Measurements for more information). In order to perform the same statistical measurement in the ensemble domain, 21 ensembles were created by picking a random sample of 11 `strong' classifiers from the pool. The agreement between the classifiers is better than moderate and, as expected, the agreement of the ensembles is high.}\label{table:summary1a}
\end{table}

\begin{table}[h!]
	\centering
	\begin{tabular}{|l|c|c|c|c|}
		\hline
		& \begin{tabular}[c]{@{}c@{}}Segmentation\\ I\end{tabular} & \begin{tabular}[c]{@{}c@{}}Segmentation\\ II\end{tabular} & \begin{tabular}[c]{@{}c@{}}Segmentation\\ III\end{tabular} & \begin{tabular}[c]{@{}c@{}}Segmentation\\ IV\end{tabular} \\ \hline
		Thigmotaxis           & 27.7\%                                                     & 24.0\%                                                      & 24.6\%                                                       & 22.5\%                                                      \\ \hline
		Incursion             & 19.0\%                                                     & 18.9\%                                                      & 20.6\%                                                       & 17.0\%                                                      \\ \hline
		Scanning              & 10.2\%                                                     & 12.3\%                                                      & 10.5\%                                                       & 11.9\%                                                      \\ \hline
		Focused Search        & 9.2\%                                                      & 8.9\%                                                       & 8.2\%                                                        & 10.0\%                                                       \\ \hline
		Chaining Response     & 4.5\%                                                      & 5.8\%                                                       & 5.5\%                                                        & 9.8\%                                                       \\ \hline
		Self Orienting        & 7.1\%                                                      & 8.8\%                                                       & 8.2\%                                                        & 8.4\%                                                       \\ \hline
		Scanning Surroundings & 17.4\%                                                     & 15.8\%                                                      & 16.8\%                                                       & 12.9\%                                                      \\ \hline
		Target Scanning       & 4.9\%                                                      & 5.6\%                                                      & 5.6\%                                                        & 7.4\%                                                       \\ \hline
		Unclassified          & 0.0\%                                                        & 0.0\%                                                      & 0.0\%                                                       & 0.1\%                                                      \\ \hline
	\end{tabular}
	\caption{\textbf{Percentage of segments falling under each class for the four segmentation configurations of Table \ref{table:summary1}.} Some differences among the four segmentations are visible although based on the results of figure \ref{fig:res_main_1} consistency on the conclusions is preserved in segmentations II, III and IV. Regarding segmentation I, where there is no indication of any difference between the two animal groups regarding the Chaining Response strategy,;more segments are identified as Thigmotaxis and Scanning Surroundings. This indicates the possibility that some segments which transit between Chaining Response and one of these strategies are classified either as Thigmotaxis or Scanning Surroundings.}\label{table:segments_count}
\end{table}

\cleardoublepage

\section{Discussion}
Methodologies that classify swimming paths in MWM to behavioural classes can reveal different stages of learning in animal groups. However, up to now, there are very few examples of earlier research that have made use of machine learning techniques to automatically detect animal behaviours. Most of them have proposed methods that are difficult to generalise and require machine learning knowledge. In our previous study \cite{gehring2015detailed} we addressed some limitations of the previous techniques by focusing on the fact that forcing whole swimming paths into a single class of behaviour can be suboptimal as each trajectory incorporates a number of different behaviours. Our methodology of detailed trajectory classification can reveal additional behavioural differences between two groups of animals and can be used even when small amount of trajectory data are available since the segmentation process, due to overlapping, typically creates a significant amount of data. Nevertheless, our previously proposed method of segmented trajectories classification required a decent amount of machine learning knowledge to be used correctly, and allowed an amount of subjectivity when choosing classifiers.

In this work, we address these issues by proposing to improve the robustness of the technique via majority voting. Our results are no longer based on a single classification tuning (classifier) but on the agreement of many. This technique alleviated the subjective assignment of the swimming path segments to classes since, in practice, many classifiers that seemingly perform equally well in validation, have relatively high disagreement, and how to best chose among them might be unclear. Here, we systematically investigate different segmentations to identify what are the bounds under which our method produce meaningful results (minimum and maximum segmentation length and number of labels that needs to be provided). Furthermore, the binomial confidence intervals on the ensemble of the classifiers are informative regarding the quality of our results.

Our methodology leads to results that differ from our earlier work \cite{gehring2015detailed}: they do not detect any significant difference for the scanning strategy. This is due to a number of factors: (i) the use of only one classifier, which results in higher error (see also figure \ref{fig:res_main_1}), (ii) the merging of three different segmentations that resulted in classifications that didn't fully agree with each other. Here we base our conclusions on the majority voting of many classifiers that are shown to have an improved performance versus the single classifiers, and therefore lead to more reliable results.

One important point that should be mentioned is that despite the fact that for each segmentation the ensemble formed has extremely low to zero error (\%), the largest segmentation failed to indicate difference on the chaining response strategy. We have identified as cause of this issue the difficulty involved when labelling large segments; in this case the chaining response can be masked by more dominant classes such as Thigmotaxis. It is worthy noting that the smoothing function, which is used to map the segments back to the whole trajectories, again do not affect the conclusions formed based on the strategies. Even without the smoothing function, again, three segmentations agree on the differences between the two animal groups on the Thigmotaxis, Incursion and Chaining Response strategies while the segmentation with the more lengthy segments cannot capture the difference on Chaining Response (refer to the appendix for the non-smoothed classification results). For this reason, the criterion for correct classification cannot be based on the classification error alone. We also require consistent results within a reasonable variation of the segmentation length, in our case 200-250cm, i.e., 2R and 2.5R, with R being the radius of the maze.

To facilitate the use of this methodology by the scientific community, we provide a complete software incorporating of our framework which includes a Graphical User Interface (GUI) to guide the user throughout all the analysis stages and allows for the manual configuration of each procedure.

Finally, it should be noted that the work we present here can generalise to other species of rodents inside the MWM (e.g. mice) as well as other experiments similar to the MWM (e.g. open field tasks, place avoidance). Two main significant changes to be made are the strategy definitions and in case of other tasks involving navigation and the trajectory features. In our recent work \cite{gehring2017analysis} we addressed the issue of pre-defined strategies by using a fully unsupervised procedure to find patterns of behaviour in the active allothetic place avoidance task. In that experiment there is no previous knowledge of animal behaviours thus supervised or semi-supervised techniques cannot be applied. However, we mentioned that our classification depends on the trajectory features that we used. A combined work of the classification boosting technique, an unsupervised methodology \cite{gehring2017analysis}, and the engineering of trajectory features that not linked to a specific experiment has the potential to lead to a robust generalised framework of trajectory analysis for many different animal species used in experimental procedures (e.g. octopus \cite{boal2000experimental} and zebrafish \cite{gerlai2017zebrafish}).

\section*{Acknowledgments}
E.V. acknowledges the grant number 264872 by EC-FP7-PEOPLE from the NAMASEN Marie-Curie Initial Training Network. C.S. acknowledges the grant number 31003A\_176206 from the Swiss National Science Foundation Project. K.L. acknowledges the grant number W19/7.PR/2014 from the Polish Ministry of Science and Education and the grant number 602102 from the FP7-HEALTH project (EPITARGET).


\section*{Conflict of Interest Statement}
The authors declare that the research was conducted in the absence of any commercial or financial relationships that could be construed as a potential conflict of interest.

\section*{Author Contributions}
A.V. and E.V. designed the methodology and wrote the manuscript. A.V. developed the stand-alone RODA software and performed the experiments. T.V.G. and E.V. designed the initial methodology. T.V.G. developed the original software tools. K.S., A.J., K.L., W.K. validated and tested the analysis and the software. M.C. contributed to the software development. C.S. provided the MWM experimental data. Z.T., tested alternative classification methods. All authors provided feedback to the manuscript.
 


\section*{References}

\bibliography{mybibfile}

\begin{thebibliography}{10}
\bibitem{gerlai2017zebrafish}
R.~Gerlai, Zebrafish and relational memory: Could a simple fish be useful for
  the analysis of biological mechanisms of complex vertebrate learning?,
  Behavioural Processes.
\end{thebibliography}
\cleardoublepage

\appendix

\section{Examples of Rodents Stereotypical Behaviours}

\begin{figure}[h!]
	\begin{center}
		\includegraphics[width=0.7\linewidth]{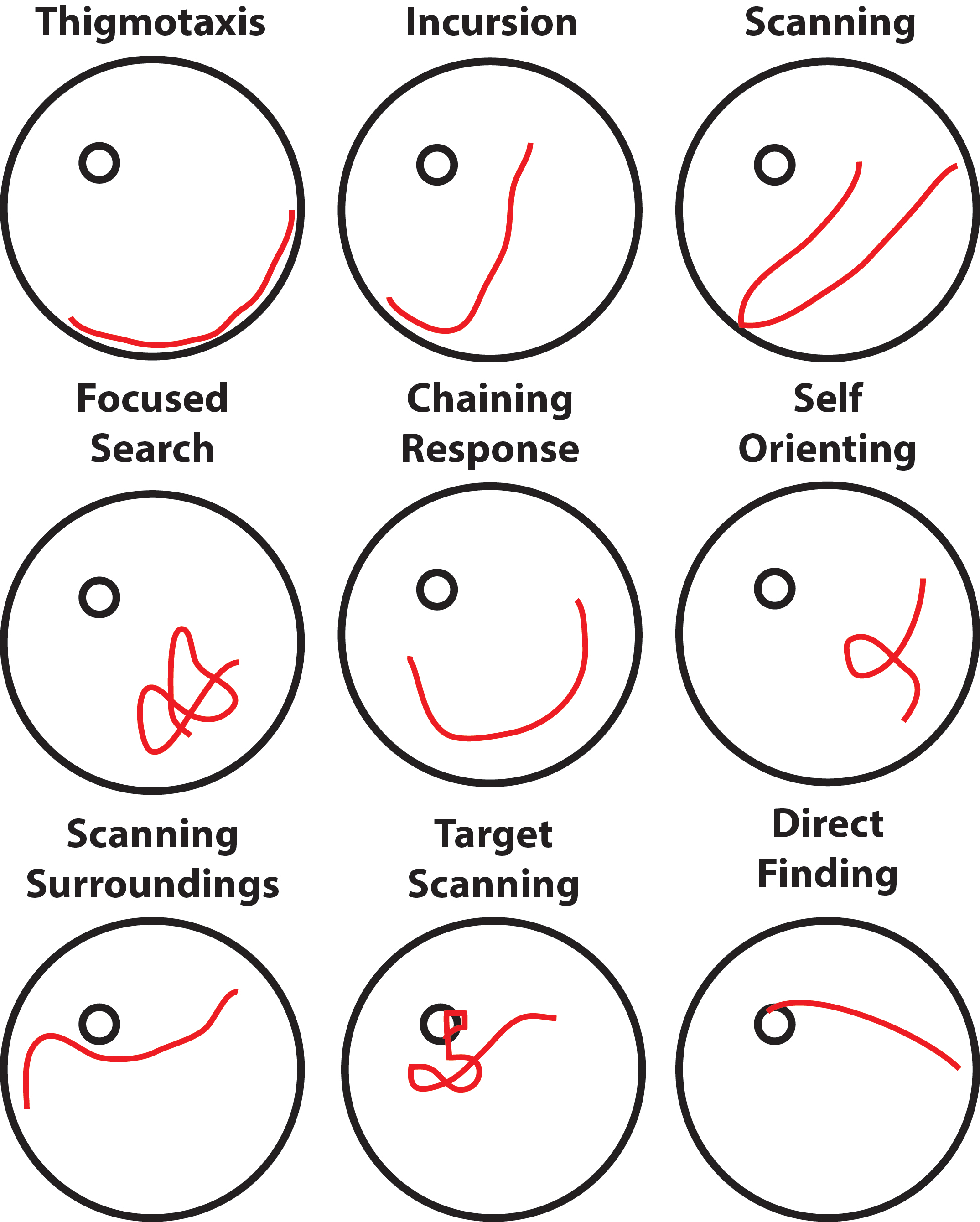}
	\end{center}
	\caption{\textbf{Stereotypical classes of behaviour.} Each figure shows an example of a trajectory segment falling under each behavioural class. Throughout the experiment, the animals implement different strategies in order to solve the maze. By detailed analysis of each trial trajectory data into segments the interchange of these stereotypical animal behaviours become visible.}\label{fig:strategies}
\end{figure}

\cleardoublepage

\section{Semi-supervised Classification Algorithm}
Our classification algorithm is based on the Metric Pairwise Constrained K-Means (MPCKMeans) clustering algorithm implemented by Bilenko et al. \cite{bilenko2004integrating}. 

MPCKMeans is inspired by the standard K-means \cite{jain2010data} clustering algorithm and belongs to the family of semi-supervised algorithms. It is able to organise a set of datapoints into groups (clusters) according to their pattern similarities and it uses a set of labelled data (predefined knowledge) in the form of ``cannot-link'' and ``must-link'' constraints in order to improve the accuracy of assigning datapoints into clusters \cite{gehring2015detailed}. Moreover, it has the ability to create clusters of different shapes and sizes by using different metrics to minimize the distance between datapoints of the same cluster and to maximize the distance between datapoints of different clusters (features weighting) \cite{bilenko2004integrating}.

With the use of labelled data it is possible not only to guide the clustering procedure (with the creation of constraints) but to also combine clusters together and form larger groups (classes) which are actually the categories of the labelled data. This mapping of clusters into classes is illustrated in our previous work \cite{gehring2015detailed} and it is done by converting a cluster into a class based on the number of labelled segments within the cluster and its size as its is shown in the equation below:

\begin{equation}
S_i \equiv \lceil s_i*l_{min,i}\rceil, \text{  where  } l_{min,i} \equiv max(s_i^{-\gamma},l_{min}) \text{  and  }
\end{equation}
\begin{equation*}
s_i \equiv \text{cluster size, } \gamma = 0.75 \text{, } l_{min} = 0.01 \text{  (or 1\%)}
\end{equation*}

\noindent Regarding the constrains a MUST-LINK constraint is generated between two datapoints with the same label and a CANNOT-LINK constraint is generated between two datapoints with a different label. Multilabelled datapoints are considered distinctive meaning that if for example a datapoint is labelled as \textit{thigmotaxis} and \textit{incursion} then a MUST-LINK constraints will be generated only with datapoints that are also labelled as \textit{thigmotaxis} and \textit{incursion}. In addition, a constrain is created only between relatively close datapoints, i.e. if the Euclidian distance between the two labelled datapoints is less that 0.25 (the features of the datapoints are normalized between [0 1]). The last rule has been implemented in our previous work \cite{gehring2015detailed} to limit the number of generated constraints which can be significantly large and create computational issues.

\begin{figure}[h!]
	\begin{center}
		\includegraphics[width=0.8\linewidth]{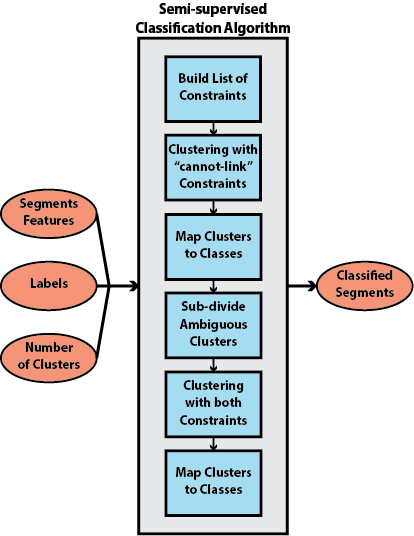}
	\end{center}
	\caption{\textbf{Stages of the Semi-supervised Classification Algorithm.} As inputs the computed features of the segments along with a partial set of labels of the segments are provided. In addition, a predefined number of target clusters needs to be provided, which specifies the number of clusters that the algorithm needs to detect. As output the algorithm provides the class in which each segment falls into. The labels are used to formed the list of constraints of which data should not be (`cannot-link') or should be (`must-link') in the same clusters. In the first-stage clustering only the `cannot-link' constrains are used to guide the clustering procedure and then clusters that could not be mapped into classes are sub-divided and a second clustering stage begins this time with both cannot-link and must-link constraints.}\label{fig:classification_algo}
\end{figure}

In order to improve the classification quality we performed the `two-stage clustering' where first we cluster the data using only the ``cannot-link'' constraints and then clusters that could not be mapped to a class are sub-divided by another clustering step, this time, however, both ``cannot-link'' and ``must-link'' constraints are used. Moreover multiple target number of clusters are tried in succession from 2 up to two times the initial number of clusters used in the first clustering. A sub-partioning is considered correctly if one of the sub-clusters could be classified. The stages of this process are shown in figure \ref{fig:classification_algo}.

\cleardoublepage

\section{10-Fold Cross Validation (for tuning and testing)}

\begin{figure}[h!]
	\begin{center}
		\includegraphics[width=\linewidth]{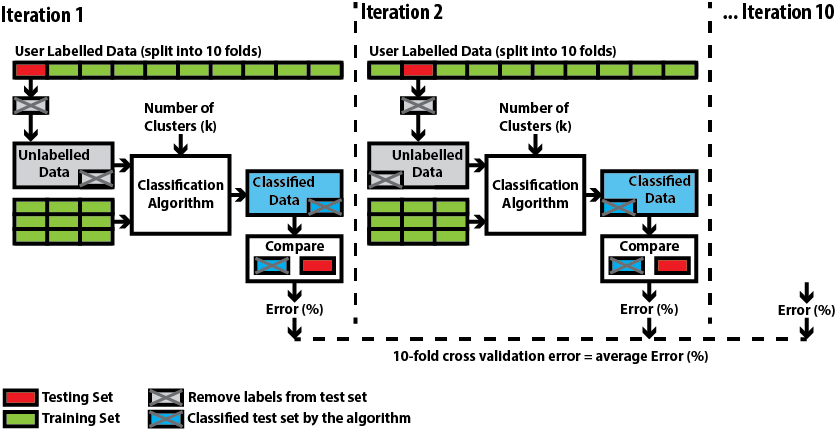}
	\end{center}
	\caption{\textbf{10-Fold cross validation process.} The manually labelled data are split into 10 folds and 9 folds are used for training and 1 fold for testing the classification results. The classification error (expressed in percentage) is calculated based on the differences between the test set and the classified test set. The same process is repeated 10 times and each time the testing set is changed with a different portion of the training set. The 10-fold cross validation error is the average error from each iteration. This process is used for the selection of the strong classifiers and for the estimation of the ensemble error. For the later, the same folds which were used to estimate the error of the classifiers are used to estimate the error of the ensemble. More strictly, in cross validation the algorithm under testing must be trained exclusively on the training set and do not receive any input from the test test (in our case the algorithm is trained using all the data but less labels). However, we use cross validation for both tuning and testing because the objective is not to create a generic classification but a specific one for the target dataset (other datasets would again require 8\% to 12\% labelling) \cite{gehring2015detailed}. The ensemble results of two segmentations (Segmentations II and IV) were manually assessed.}
	\label{fig:10foldCV}
\end{figure}

\begin{table}[h!]
	\centering
	\label{manuallyT}
	\begin{tabular}{|c|c|c|}
		\hline
		& Segmentation II & Segmentation IV \\ \hline
		TT      & 1.4\%           & 1.0\%                \\ \hline
		IC      & 8.0\%           & 1.5\%                \\ \hline
		SC      & 6.6\%           & 4.1\%                \\ \hline
		FS      & 6.0\%           & 3.7\%                \\ \hline
		CR      & 11.5\%          & 12.6\%                \\ \hline
		SO      & 11.0\%          & 8.1\%                \\ \hline
		SS      & 9.6\%           & 3.2\%                \\ \hline
		ST      & 2.3\%           & 1.0\%                \\ \hline
		average & 5.7\%           & 4.4\%                \\ \hline
		total   & 6.3\%           & 2.8\%                \\ \hline
	\end{tabular}
	\caption{\textbf{Manual estimation of ensemble error.} The ensemble error was manually assessed for the segmentations II and IV. The table shows both the total error and the error among the different classes (including the average). The total manually estimated error of the ensembles is still significant lower than the average error of the classifiers (6.3\% vs 17.5\% for Segmentation II and 2.8\% vs 18.0\% for Segmentation IV). The results of the ensembles were manually assessed for two reasons: (i) to estimate the overfitting, which is likely to be caused because the same data for both tuning and testing, and (ii) because our testing set was very small since limited amount of labels have been provided, the error estimation is likely to be overly optimistic.}
\end{table}

\cleardoublepage

\section{Smoothing Function}

\begin{figure}[h!]
	\begin{center}
		\includegraphics[width=0.5\linewidth]{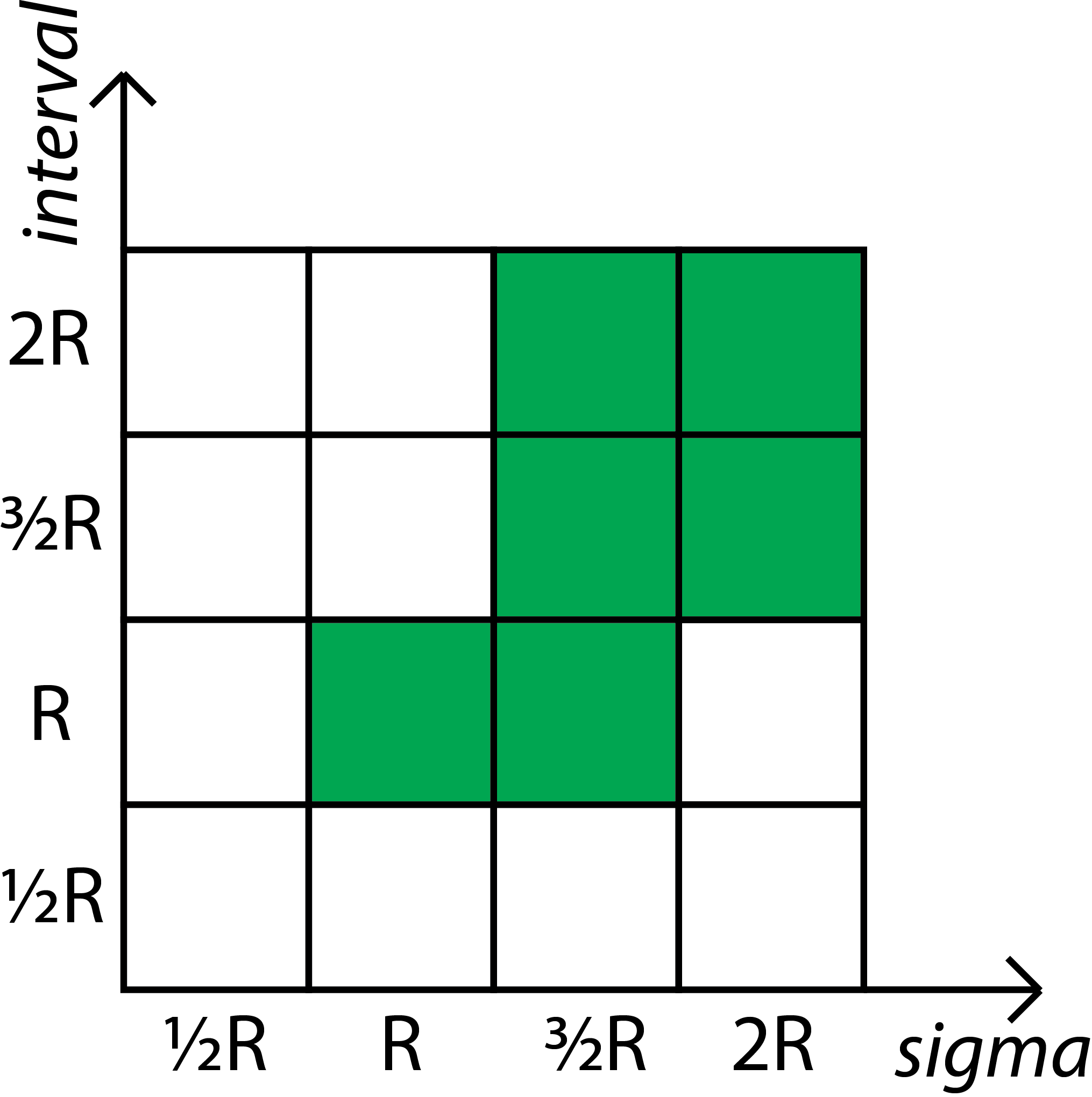}
	\end{center}
	\caption{\textbf{Empirically defined area of tuning for the smoothing function.} R refers to the arena radius (in cm); x-axis (sigma) refers to a particular value of $\sigma$ (variance of the Gaussian); y-axis (interval) refers to a particular value of the length of the interval; green boxes indicate areas under which the smoothing function (refer to section 2.7 Mapping Segment Classes to the Full Swimming Paths) yields consistent results for every segmentation (excluding Segmentation I where the segments length is too large). Interval of length $2 \cdot R$ is at the limit and from this point onwards consistency cannot be sustained.}\label{fig:smoothing}
\end{figure}

\cleardoublepage

\section{Agreement Matrix}

\begin{figure}[h!]
	\begin{center}
		\includegraphics[width=\linewidth]{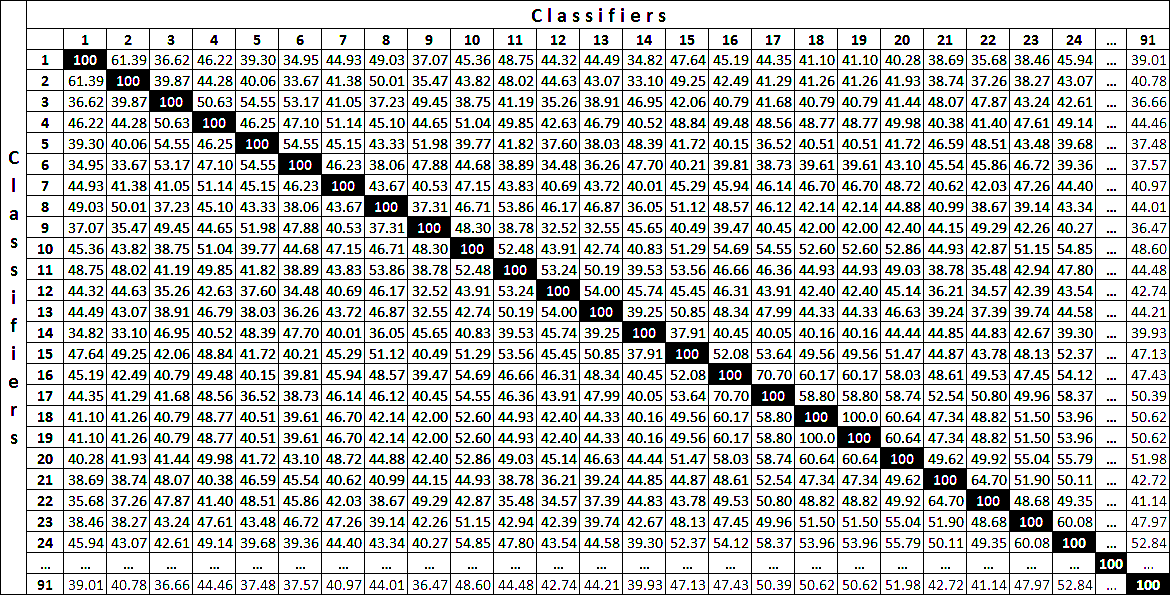}
	\end{center}
	\caption{\textbf{Agreement matrix for the classifiers of Segmentation III (see Table \ref{table:summary1}).} The classifier of each column is being compared with the classifier of each row. The comparison is based on the percentage of segments that both classifiers agree that belong to the same class. The diagonal values of the matrix indicate 100\% agreement since each classifier is compared with itself.}\label{fig:agreement_matrix}
\end{figure}

\cleardoublepage

\section{Results of each Segmentation without the Smoothing Function}
\begin{figure}[h!]
	\begin{center}
		\includegraphics[width=0.9\linewidth]{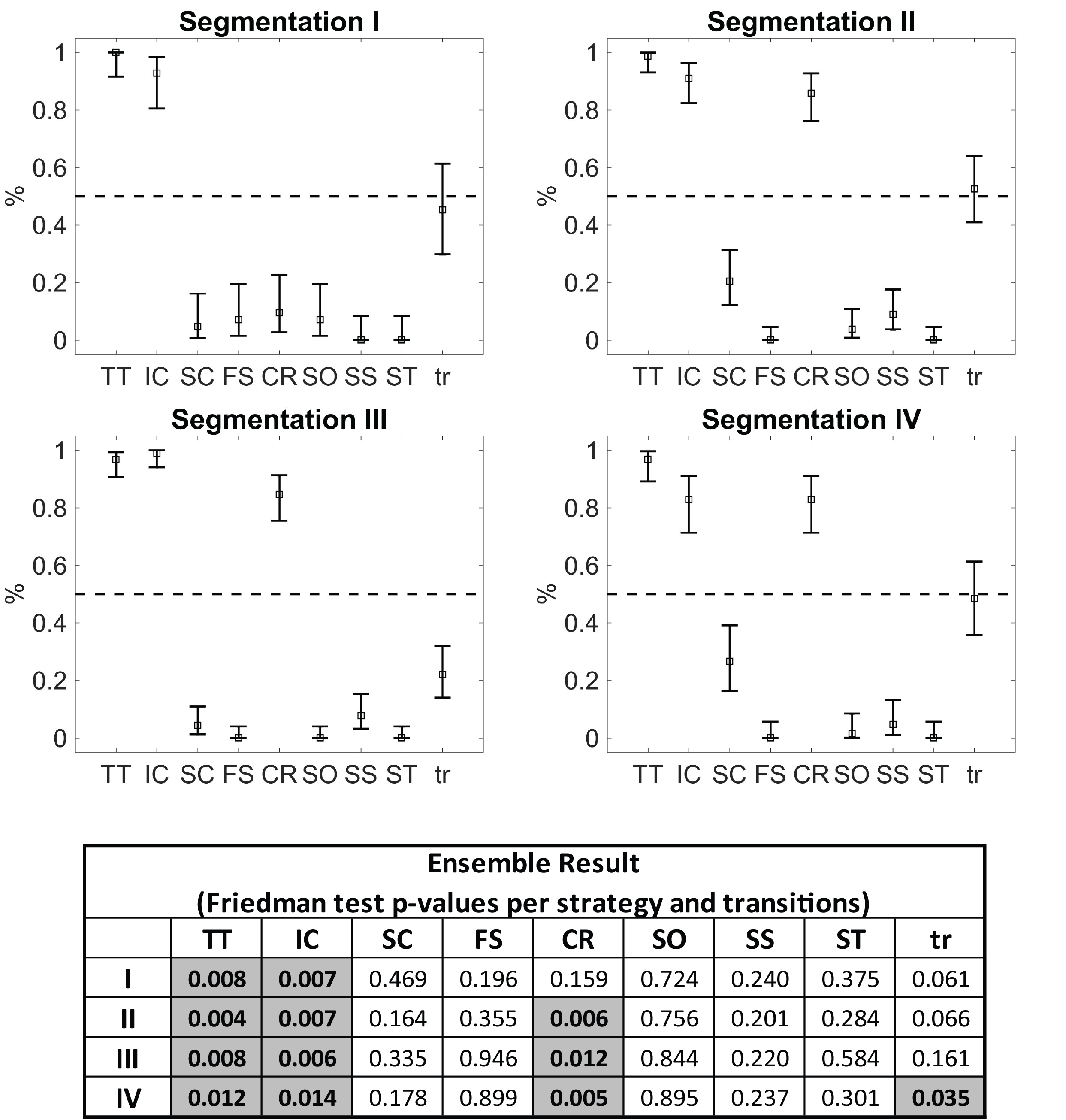}
	\end{center}
	\caption{\textbf{Conclusive pre-smoothing results from the classification of each segmentation (see Table \ref{table:summary1}).} Considering segments as continuous parts of the trajectories ignoring the overlapping provides consistent results when differences between the implemented strategies of the groups are being investigated but creates an overestimation on the number of transitions between strategies. Each plot shows the 95\% binomial confidence intervals for the classifiers of each segmentation regarding their agreement if there is significant difference between the two animal groups (Friedman test p-value $< 0.05$) on each strategy and strategy transitions or not. Squares indicate the mean of the classifiers that point out that there is significant difference on this particular case; errorbars are the 95\% confidence intervals; the dashed line indicates the threshold of interest (0.5 or 50\%). In order to be confident that there is indeed a significant difference between the the two animal groups on each strategy and the strategy transitions the confidence intervals should be clearly above 0.5 (or 50\%). Compared to the results in the main manuscript (refer to figure \ref{fig:res_main_1}) we see that the smoothing function which maps the segments to the full swimming paths is actually beneficial on revealing the animal transitions between strategies. Other than that the results and the conclusions are the same.}\label{fig:no-smoothing-class}
\end{figure}

\begin{table}[h!]
	\centering
	\begin{tabular}{|l|c|c|c|c|}
		\hline
		\multicolumn{1}{|c|}{}                                               & \begin{tabular}[c]{@{}c@{}}Segmentation\\ I\end{tabular}            & \begin{tabular}[c]{@{}c@{}}Segmentation\\ II\end{tabular}           & \begin{tabular}[c]{@{}c@{}}Segmentation\\ III\end{tabular}          & \begin{tabular}[c]{@{}c@{}}Segmentation\\ IV\end{tabular}           \\ \hline
		& \multicolumn{4}{c|}{Classifiers}                                                                                                                                                                                                                                                      \\ \hline
		\begin{tabular}[c]{@{}l@{}}Unclassified (\%)\\ Segments\end{tabular} & 24.8                                                              & 24.3                                                               & 30.0                                                               & 29.0                                                             \\ \hline
		Agreement (\%)                                                       & 53.2                                                              & 55.5                                                              & 48.8                                                             & 52.1                                                            \\ \hline
		& \multicolumn{4}{c|}{Ensemble(s)}                                                                                                                                                                                                                                                      \\ \hline
		\begin{tabular}[c]{@{}l@{}}Unclassified (\%)\\ Segments\end{tabular} & 1.2                                                              & 0.7                                                           & 0.8                                                              & 1.1                                                              \\ \hline
		Agreement (\%)                                                       & 84.7                                                             & 83.3                                                              & 79.8                                                              & 80.0                                                              \\ \hline
	\end{tabular}
	\caption{\textbf{Classification statistics for the four segmentation configurations of Table \ref{table:summary1} prior to smoothing.} In comparison with the results of Table \ref{table:summary1a} we see that the percentage of unclassified segments among the classifiers is higher and the agreement between them lower. However, the ensemble (or ensembles in case of the agreement) again nearly nullifies the unclassified segments and significantly boosts the agreement percentage.}\label{table:presummary1a}
\end{table}

\cleardoublepage

\section{Ensemble Results of each Segmentation}
\begin{figure}[h!]
	\centering
	\includegraphics[width=\linewidth]{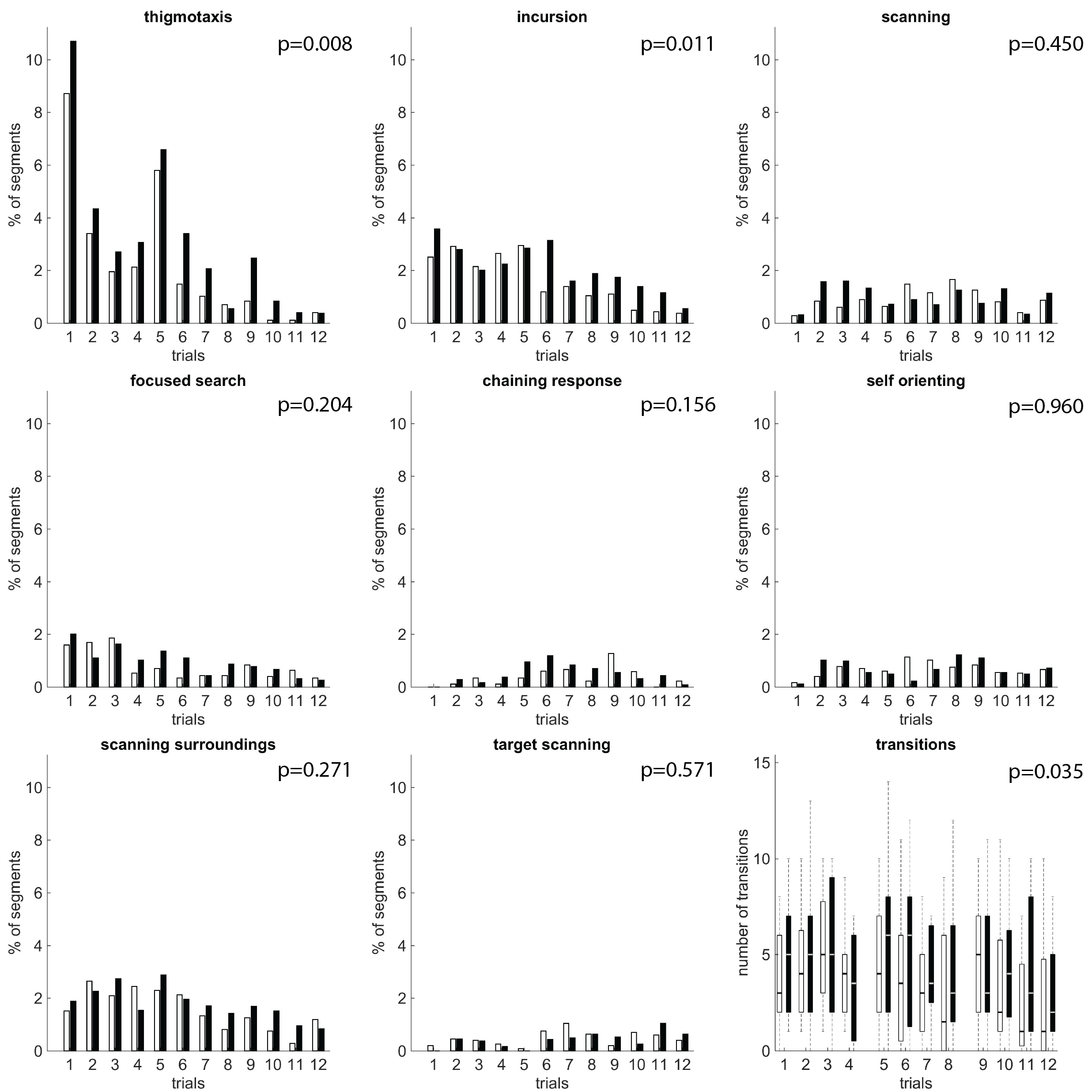}
	\caption{\textbf{Percentage of segments falling under each strategy for the stress (black) and control (white) animal groups over each trial for the Segmentation I of Table \ref{table:summary1}.} All the animals were tested for a set of 12 trials divided in 3 sessions (days). Each segment is considered to be of length equal to the length of the arena radius (100cm). For the transitions: bars represent the first and third quartiles of the data; the black (control group) or white (stress group) horizontal lines is the median, crosses are the outliers and whiskers indicate the minimum and the maximum values. The Friedman test p-value (shown on the top right) was used to compare both animal groups for the complete set of trials. According to the plots stress animals swap longer paths since the average number of strategy implementations in higher than the one of the control group. Thigmotaxis and Incursion strategies show a clear difference in favor of the stress groups along with the strategies transitions. This Segmentation configuration fails to reveal significant different on the Chaining Response because of the segment length which causes some more rare strategies to disappear.}\label{extra1}
\end{figure}

\begin{figure}[h!]
	\centering
	\includegraphics[width=\linewidth]{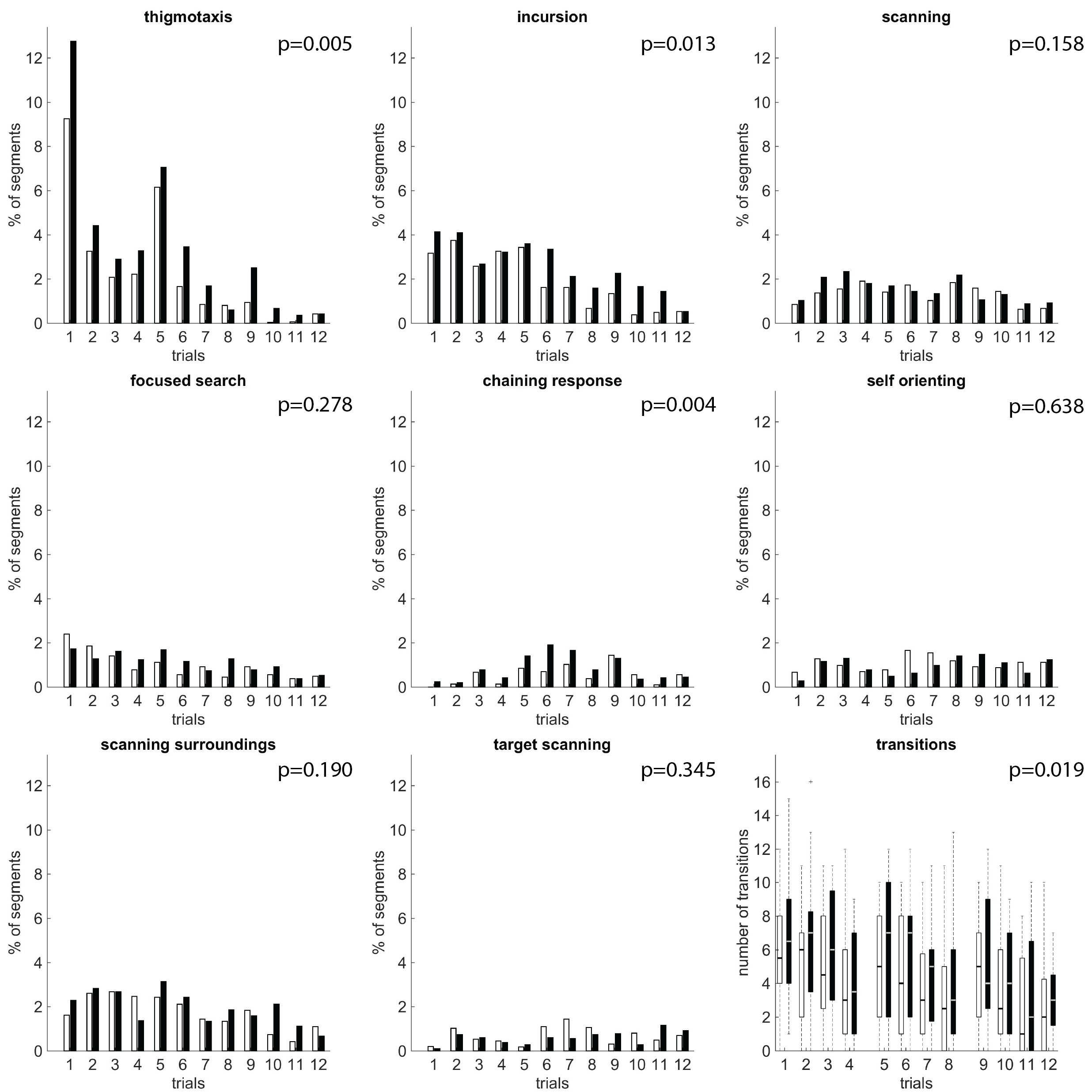}
	\caption{\textbf{Percentage of segments falling under each strategy for the stress (black) and control (white) animal groups over each trial for the Segmentation II of Table \ref{table:summary1}.} All the animals were tested for a set of 12 trials divided in 3 sessions (days). Each segment is considered to be of length equal to the length of the arena radius (100cm). For the transitions: bars represent the first and third quartiles of the data; the black (control group) or white (stress group) horizontal lines is the median, crosses are the outliers and whiskers indicate the minimum and the maximum values. The Friedman test p-value (shown on the top right) was used to compare both animal groups for the complete set of trials. According to the plots stress animals swap longer paths since the average number of strategy implementations in higher than the one of the control group. Thigmotaxis and Incursion strategies show a clear difference in favor of the stress groups along with Chaining Response, which is not implemented systematically though. The number of transitions between strategies shows that the stress animals change their behaviour more ofter within single trials.}\label{extra2}
\end{figure}

\begin{figure}[h]
	\centering
	\includegraphics[width=\linewidth]{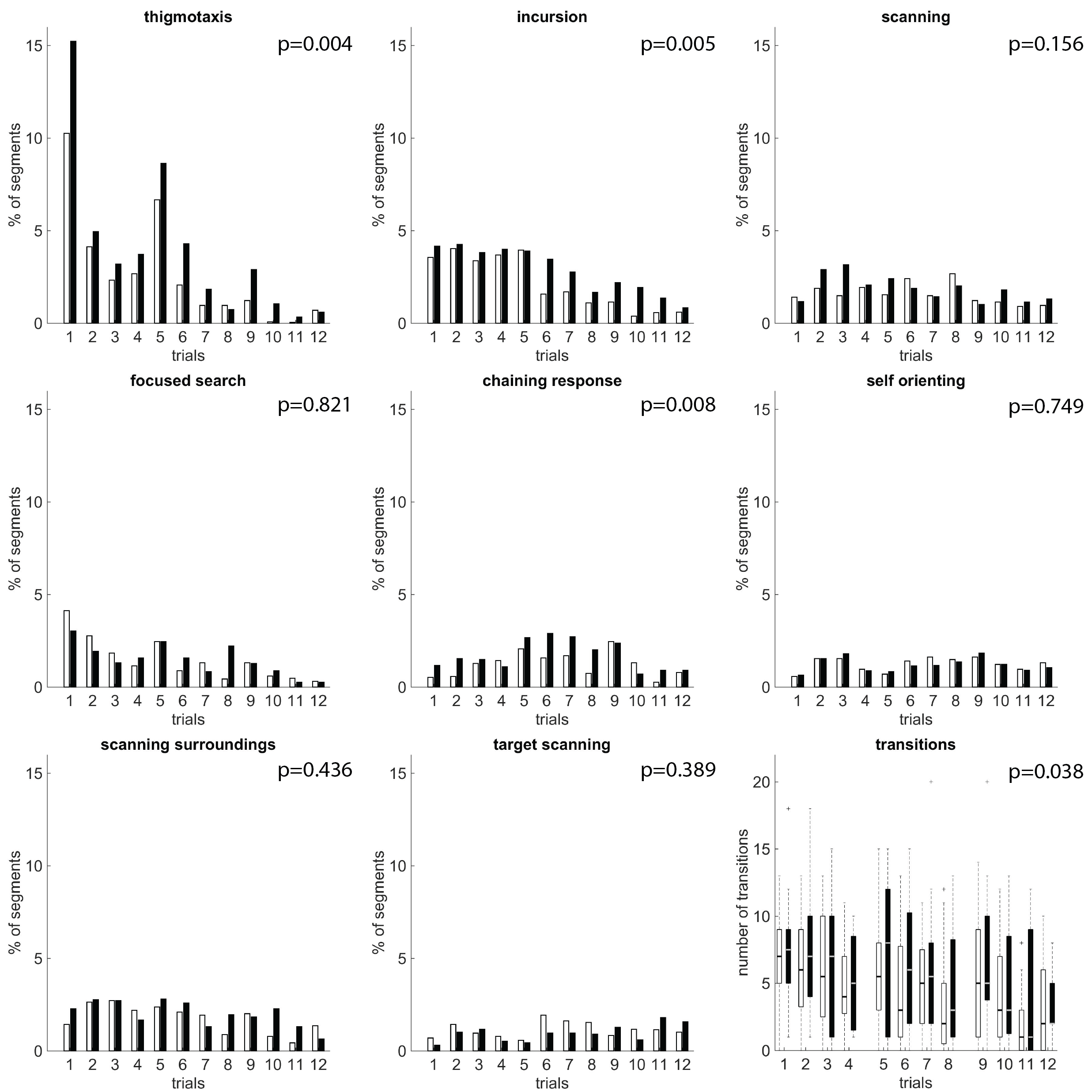}
	\caption{\textbf{Percentage of segments falling under each strategy for the stress (black) and control (white) animal groups over each trial for the Segmentation IV of Table \ref{table:summary1}.} All the animals were tested for a set of 12 trials divided in 3 sessions (days). Each segment is considered to be of length equal to the length of the arena radius (100cm). For the transitions: bars represent the first and third quartiles of the data; the black (control group) or white (stress group) horizontal lines is the median, crosses are the outliers and whiskers indicate the minimum and the maximum values. The Friedman test p-value (shown on the top right) was used to compare both animal groups for the complete set of trials. According to the plots stress animals swap longer paths since the average number of strategy implementations in higher than the one of the control group. Thigmotaxis and Incursion strategies show a clear difference in favor of the stress groups along with Chaining Response, which is not implemented systematically though. The number of transitions between strategies shows that the stress animals change their behaviour more ofter within single trials.}\label{extra4}
\end{figure}

\cleardoublepage

\end{document}